\documentclass[aps,prb,twocolumn,superscriptaddress,floatfix,longbibliography]{revtex4-1}
\usepackage{lineno}
\usepackage{pdftexcmds}
\usepackage{graphicx}  
\usepackage{dcolumn}   
\usepackage{bm}        
\usepackage{amssymb}   
\usepackage{amsmath}
\usepackage{blkarray, multirow, graphicx, diagbox, color, colortbl}
\usepackage{bbm, bbold}
\usepackage{ifthen}

\usepackage{xkeyval}
\usepackage{moreverb}
\usepackage{rotating}
\usepackage{slashbox}
\usepackage{xspace}
\usepackage{capt-of}
\usepackage[caption=false]{subfig}
\newcommand\subfigref[1]{\protect\subref{#1}}

\usepackage[svgnames]{xcolor}

\usepackage{glossaries}
\glsdisablehyper 
\usepackage{hyphenat}
\usepackage{nicefrac}
\usepackage[]{units}
\usepackage{physics}
\usepackage{braket}
\usepackage[inline]{enumitem}
\usepackage{tabto}
\usepackage{listings}
\usepackage{xstring}
\def\ReplaceStr#1{%
	\IfSubStr{#1}{p}{%
		\StrSubstitute{#1}{p}{.}}{#1}}
\def\ReReplaceStr#1{%
	\IfSubStr{#1}{.}{%
		\StrSubstitute{#1}{.}{p}}{#1}}
\usepackage{tikz}
\usepackage{calc}
\usepackage{pgffor}
\usepackage{pgfplots}
\pgfplotsset{compat=1.13}
\usepackage{pgfplotstable}
\usepgfplotslibrary{groupplots}
\usepgfplotslibrary{fillbetween}

\tikzstyle{n} = [draw,shape=ellipse,minimum size=1.5em,inner sep=0pt,fill=white!20, minimum width=2.5em]
\tikzstyle{Init} = [n,color=green,fill=green!20,text=black]
\tikzstyle{Fin} = [n,color=red,fill=red!20,text=black]
\tikzstyle{Ghost} = [minimum size=1.5em,inner sep=0pt,color=white,text=black]
\tikzstyle{Multiple} = [draw,shape=rect,minimum size=2em,inner sep=0pt]

\tikzstyle{ghostA} = [text=red!70,thick, minimum size=2*(5pt-\pgflinewidth), inner sep=0pt, outer sep=0pt]
\tikzstyle{ghostB} = [text=blue!70,thick, minimum size=2*(5pt-\pgflinewidth), inner sep=0pt, outer sep=0pt]
\tikzstyle{siteA} = [draw=red!70,circle,thick, minimum size=2*(5pt-\pgflinewidth), inner sep=0pt, outer sep=0pt]
\tikzstyle{siteB} = [draw=blue!70,circle,thick, minimum size=2*(5pt-\pgflinewidth), inner sep=0pt, outer sep=0pt]
\tikzstyle{operatorA} = [cross out, draw=red!70, thick, minimum size=2*(5pt-\pgflinewidth), inner sep=0pt, outer sep=0pt]
\tikzstyle{operatorB} = [cross out, draw=blue!70, thick, minimum size=2*(5pt-\pgflinewidth), inner sep=0pt, outer sep=0pt]

\definecolor{colorA}{rgb} {0.58,0,0.8275}
\definecolor{colorB}{rgb} {0.11,0.663,0.51}
\definecolor{colorC}{rgb} {0.3373,0.7059,0.9137}
\definecolor{colorD}{rgb} {0.902,0.6235,0}
\definecolor{colorE}{rgb} {0.9451,0.902,0.3255}
\definecolor{colorIPphase}{rgb} {0.4, 0.7, 0.4}
\definecolor{colorIPcharge}{rgb} {0.7, 0.1, 0.7}
\definecolor{shinyblue}{HTML} {257eee}
\definecolor{fatyellow}{HTML} {fbfe36}
\definecolor{glowyred}{HTML} {ea1822}
\definecolor{midnight_blue}{HTML} {002952}
\definecolor{turquoiseblue}{rgb}{0.0, 1.0, 0.94}
\definecolor{redviolet}{HTML}{A1246B}

\usetikzlibrary
{
	calc,
	decorations,
	fadings,
	plotmarks,
	patterns,
	positioning,
	petri,
	arrows,
	decorations.markings,
	backgrounds,
	fit,
	graphs,
	shapes.geometric,
	decorations.pathmorphing,
	shapes.misc,
	shapes,
	tikzmark,
	pgfplots.colorbrewer,
	fpu
}

\pgfplotsset{
        cycle from colormap manual style/.style={
            x=3cm,y=10pt,ytick=\empty,
            stack plots=y,
            every axis plot/.style={line width=2pt},
        },
}

\tikzset
{
	style matrix_highlight_green/.style=
	{
		set fill color=green!20,
		set border color=green!50,
		fill opacity=0.5
	},
	style matrix_highlight_red/.style=
	{
		set fill color=red!20,
		set border color=red!50,
		fill opacity=0.5
	},
	style matrix_highlight_blue/.style=
	{
		set fill color=cyan!20,
		set border color=blue!40
	},
	hor/.style=
	{
		above left offset={-0.15,0.5},
		below right offset={0.15,-0.15},
		#1
	},
	hor2/.style=
	{
		above left offset={-0.15,0.4},
		below right offset={0.15,-0.15},
		#1
	},
	hor3/.style=
	{
		above left offset={-0.15,0.25},
		below right offset={0.15,0.0},
		#1
	},
	ver/.style=
	{
		above left offset={-0.1,0.5},
		below right offset={0.15,-0.15},
		#1
	}
}

\tikzset{>=stealth}
\tikzset{->-/.style={decoration={
			markings,
			mark=at position .5 with {\arrow{>}}},postaction={decorate}}}

\tikzstyle{orientedsnake} = [
decorate, 
decoration={snake},
->
]  
\tikzstyle{orientedshortarrow} = [
decoration={markings,
	mark=at position .33 with {\arrow{>}}},
postaction={decorate}
]  
\tikzstyle{orientedlongarrow} = [
decoration={markings,
	mark=at position .67 with {\arrow{>}}},
postaction={decorate}
]
\tikzset{dbl/.style={double,
		double equal sign distance,
		-implies,
		shorten >=10pt,
		shorten <=10pt}}
\tikzset{
	between/.style args={#1 and #2}{
		at = ($(#1)!0.5!(#2)$)
	}
}
\pgfmathdeclarefunction{quadFct}{3}%
{%
	\pgfmathparse{#1*x*x+#2*x+#3}%
}
\pgfmathdeclarefunction{linearFct}{2}%
{%
	\pgfmathparse{#1*x+#2}%
}
\pgfmathdeclarefunction{logFct}{2}%
{%
	\pgfmathparse{#1*log10(x)+#2}%
}
\pgfmathdeclarefunction{expFct}{2}%
{%
	\pgfmathparse{#1*exp(-x/#2)}%
}
\pgfmathdeclarefunction{ccFct}{3}
{%
	\pgfmathparse{(#1)/6.0*ln((#2)*sin(deg((pi*x)/(#2))))+(#3)}%
}%
\pgfmathdeclarefunction{lorentzian}{2}%
{%
	\pgfmathparse{1.0/(pi*#2*(1.0+((x-#1)/(#2))^2))}%
}%

\newcommand{\nodagger}[0]{{\phantom{\dagger}}}
\newcommand{\noprime}[0]{{\phantom{\prime}}}

\pgfmathdeclarefunction{peierlspotential}{6}%
{
	\pgfmathparse
	{
		#2 / (#3 * #5) 
		* sin(deg(#1 * #3 - #3 * #5 * (x)))
		* exp(-( ( #1 - #5 * ((x) - #4) ) )^2.0 / ( #6^2.0 ) )
	}%
}
\newboolean{buildCurrentBlockInline}
\setboolean{buildCurrentBlockInline}{false}
\newboolean{rebuildCurrentBlock}
\setboolean{rebuildCurrentBlock}{false}

\newboolean{rebuildData}
\setboolean{rebuildData}{true}

\newboolean{buildtikzpics}
\setboolean{buildtikzpics}{true}

\newboolean{useShellEscape}
\setboolean{useShellEscape}{true}

\pgfdeclareverticalshading{bwGood}{100bp}
{
	color(0bp)=(white); 
	color(25bp)=(white); 
	color(45bp)=(gray);
	color(49bp)=(black); 
	color(51bp)=(black); 
	color(55bp)=(gray);
	color(75bp)=(white); 
	color(100bp)=(white)
}

\pgfdeclareverticalshading{bwBad}{100bp}
{
	color(0bp)=(white); 
	color(25bp)=(white); 
	color(45bp)=(red!40);
	color(49bp)=(red!80!black); 
	color(51bp)=(red!80!black); 
	color(55bp)=(red!40);
	color(75bp)=(white); 
	color(100bp)=(white)
}

\newif\ifrebuildtikz
\newif\ifChangeMode
\ChangeModetrue
\ChangeModefalse
\usetikzlibrary{external}
\ifthenelse{\boolean{useShellEscape}}
{
	\rebuildtikztrue
}
{
	\rebuildtikzfalse
	\tikzset{external/export=false}
}

\usepackage{soul}
\usepackage{cleveref}
\Crefname{appendix}{Appendix}{Appendices}
\Crefname{equation}{Equation}{Equations}
\Crefname{figure}{Figure}{Figures}
\Crefname{section}{Section}{Sections}
\Crefname{paragraph}{Paragraph}{Paragraphs}
\Crefname{tabular}{Tabular}{Tabulars}
\crefname{appendix}{App.}{Apps.}
\crefname{equation}{Eq.}{Eqs.}
\crefname{figure}{Fig.}{Figs.}
\crefname{section}{Sec.}{Secs.}
\crefname{paragraph}{Par.}{Pars.}
\crefname{tabular}{Tab.}{Tabs.}

\ExplSyntaxOn
\DeclareExpandableDocumentCommand \eval { m } { \fp_eval:n { #1 } }
\ExplSyntaxOff

\newcommand{\printpgfnumberwitherror}[2]%
{%
	\pgfmathfloatparsenumber{#1}%
	\pgfmathfloattomacro{\pgfmathresult}{\Fn}{\Mn}{\En}%
	\pgfmathparse{\Fn==2 ? "-" : ""}%
	\edef\Sn{\pgfmathresult}%
	\pgfmathfloatparsenumber{#2}%
	\pgfmathfloattomacro{\pgfmathresult}{\Fe}{\Me}{\Ee}%
	\pgfmathparse{int(sqrt((\Ee-\En)^2))}%
	\edef\precisionAbsEe{\pgfmathresult}%
	\pgfmathparse{int(\Ee-\En)}%
	\edef\precisionE{\pgfmathresult}%
	\pgfmathparse{\eval{\Me*10^(\precisionE)}}%
	\ifthenelse{\En=0}%
	{%
		$\Sn\pgfmathprintnumber[fixed, precision=\precisionAbsEe, zerofill]{\Mn} \pm (\pgfmathprintnumber[std, precision=0, zerofill]{#2})$%
	}%
	{%
		\ifthenelse{\En=-1}%
		{%
			\pgfmathparse{\eval{\Mn/10}}%
			\edef\Mn{\pgfmathresult}%
			\pgfmathparse{int(\precisionAbsEe+1)}%
			\edef\precisionAbsEe{\pgfmathresult}%
			$\Sn\pgfmathprintnumber[fixed, precision=\precisionAbsEe, zerofill]{\Mn} \pm (\pgfmathprintnumber[std, precision=0, zerofill]{#2})$%
		}%
		{%
			$\left(\Sn\pgfmathprintnumber[std, precision=\precisionAbsEe, zerofill]{\Mn} \pm \pgfmathprintnumber[fixed, precision=\precisionAbsEe, zerofill]{\pgfmathresult}\right)\cdot10^{\En}$%
		}%
	}%
}

\def\imgpath{.}

\newacronym{PCMO}{PCMO}{praseodymium\hyp calcium\hyp manganite}
\newacronym{1D}{1D}{one\hyp dimensional}
\newacronym{2D}{2D}{two\hyp dimensional}
\newacronym{MPS}{MPS}{matrix\hyp product state}
\newacronym{MPO}{MPO}{matrix\hyp product operator}
\newacronym{SVD}{SVD}{singular\hyp value decomposition}
\newacronym{QCS}{QCS}{quantum\hyp computer simulator}
\newacronym{QC}{QC}{quantum computer}
\newacronym{FSM}{FSM}{finite\hyp state machine}
\newacronym{ACA}{ACA}{adaptive cross\hyp approximation}
\newacronym{CDW}{CDW}{charge\hyp density wave}
\newacronym{SDW}{SDW}{spin\hyp density wave}
\newacronym{ARPES}{ARPES}{angle\hyp resolved photoemission spectroscopy}
\newacronym{OBC}{OBC}{open\hyp boundary conditions}
\newacronym{PBC}{PBC}{periodic\hyp boundary conditions}
\newacronym{TEBD}{TEBD}{time\hyp evolution block\hyp decimation}
\newacronym{iff}{iff}{if and only if}
\newacronym{DFT}{DFT}{density\hyp functional theory}
\newacronym{DMFT}{DMFT}{dynamical mean\hyp field theory}
\newacronym{CDMFT}{C-DMFT}{cluster dynamical mean\hyp field theory}
\newacronym{DMRG}{DMRG}{density\hyp matrix renormalization group}
\newacronym{QMC}{QMC}{quantum Monte Carlo}
\newacronym{AFQMC}{AFQMC}{auxiliary-field quantum Monte Carlo}
\newacronym{VCA}{VCA}{variational cluster approximation}
\newacronym{SEF}{SEF}{self\hyp energy\hyp functional}
\newacronym{AIM}{AIM}{Anderson impurity model}
\newacronym{SIAM}{SIAM}{single impurity Anderson model}
\newacronym{LDA}{LDA}{local\hyp density approximation}
\newacronym{LBNL}{LBNL}{Lawrence Berkeley National Laboratory}
\newacronym{VQE}{VQE}{variational\hyp quantum eigensolver}
\newacronym{ED}{ED}{exact diagonalization}
\newacronym{QPT}{QPT}{quantum phase transition}
\newacronym{QCP}{QCP}{quantum critical point}
\newacronym{ETH}{ETH}{eigenstate thermalization hypothesis}
\newacronym{AKLT}{AKLT}{Affleck\hyp Lieb\hyp Kennedy\hyp Tasaki}
\newglossaryentry{QR}{name={QR},description={QR decomposition}}
\newacronym{TNS}{TNS}{tensor\hyp network state}
\newacronym{NN}{NN}{nearest\hyp neighbour}
\newacronym{NNN}{NNN}{next\hyp nearest\hyp neighbour}
\newacronym{TLL}{TLL}{Tomonaga-Luttinger liquid}
\newacronym{PLL}{PLL}{pair Luttinger liquid}
\newacronym{FB}{FB}{Flat band}
\newacronym{C}{C}{coexistence}
\newacronym{SU}{SU}{superfluid}
\newacronym{PH}{PH}{projected Hamiltonian}
\newacronym{IP}{IP}{itinerant paired}
\newacronym{QO}{QO}{quasi-order}
\newacronym{PRM}{PRM}{purely repulsive model}
\newacronym{MIM}{MIM}{mixed-interaction model}
\newacronym{LL}{LL}{Luttinger liquid}
\newacronym{QET}{QET}{quantum\hyp embedding techniques}
\newacronym{AFM}{AFM}{antiferromagnetism}
\newacronym{PM}{PM}{paramagnetic}
\newacronym{MLWF}{MLWF}{maximally\hyp localized Wannier function}
\newacronym{cRPA}{cRPA}{constrained Random Phase Approximation}
\newacronym{CPT}{CPT}{cluster\hyp  perturbation theory}
\newacronym{CET}{CET}{cluster\hyp embedding technique}
\newacronym{PEPS}{PEPS}{projected entangled\hyp pair states}
\newacronym{DOS}{DOS}{density of states}
\begin{document}
\def\thetitle{Matrix product state based band Lanczos solver for quantum cluster approaches}
\title{\thetitle}
\author{Sebastian Paeckel}
\email{sebastian.paeckel@physik.uni-muenchen.de}
\thanks{These authors contributed equally.}
\affiliation{Department of Physics, Arnold Sommerfeld Center for Theoretical Physics (ASC),	Munich Center for Quantum Science and Technology (MCQST),Ludwig-Maximilians-Universität München, 80333 München, Germany}

\author{Thomas Köhler}
\email{t.kohler@hw.ac.uk}
\thanks{These authors contributed equally.}
\affiliation{Department of Physics and Astronomy, Uppsala University, Box 516, S-751 20 Uppsala, Sweden}
\affiliation{SUPA, Institute of Photonics and Quantum Sciences, Heriot-Watt University, Edinburgh EH14 4AS, United Kingdom}

\author{Salvatore R. Manmana}
\email{salvatore.manmana@theorie.physik.uni-goettingen.de}
\affiliation{Institute for Theoretical Physics, Georg-August-University G\"ottingen, 37077 G\"ottingen, Germany}

\author{Benjamin Lenz}
\email{benjamin.lenz@sorbonne-universite.fr}
\affiliation{IMPMC, Sorbonne Université, CNRS, MNHN, IRD, 4 place Jussieu, 75005 Paris, France}

\date{\today}

\begin{abstract}
	We present a \gls{MPS} based band Lanczos method as solver for quantum cluster methods such as the \gls{VCA}.
	While a na\"ive implementation of \gls{MPS} as cluster solver would barely improve its range of applicability, we show that our approach makes it possible to treat cluster geometries well beyond the reach of exact diagonalization methods.
	The key modifications we introduce are a continuous energy truncation combined with a convergence criterion that is more robust against approximation errors introduced by the \gls{MPS} representation and provides a bound to deviations in the resulting Green's function.
	The potential of the resulting cluster solver is demonstrated by computing the self\hyp energy functional for the single\hyp band Hubbard model at half filling in the strongly correlated regime, on different cluster geometries.
	Here, we find that only when treating large cluster sizes, observables can be extrapolated to the thermodynamic limit, which we demonstrate at the example of the staggered magnetization.
	Treating clusters sizes with up to $6\times 6$ sites we obtain significant improvement over the extrapolation accessible with exact diagonalization solvers when comparing to quantum Monte Carlo results.
	Finally, we illustrate the applicability of the \gls{MPS} cluster solver to more complex models by calculating spectral properties as relevant for the electron\hyp doped cuprate CaCuO$_2$.
\end{abstract}

\maketitle 

\section{Introduction}
Exploring the quantum states of matter of two\hyp dimensional strongly correlated electron systems is inherently difficult.
Prime examples are the cuprate high\hyp temperature superconductors \cite{highTc_original}, whose salient features are frequently described within a single\hyp band Hubbard\hyp type model on a square lattice geometry \cite{dagotto,PhysRevX.5.041041,2DHubbard_Science}.
More recently, two-dimensional materials have attracted substantial interest in a variety of systems, like bilayer graphene and other multi\hyp layered materials \cite{SCgraphene1,SCgraphene2,SCgraphene3}, transfer metal dichalcogenides (TMDCs) \cite{reviewTMDCs}, or kagom\'e metals \cite{kagome_metals,kagome_metals_PRL}.
Experiments on these materials have revealed unconventional phases of matter, e.g., a Wigner crystal quantum Hall phase \cite{Wigner_quantumHall}, or interesting superconducting phases \cite{SCgraphene1,SCgraphene2,SCgraphene3,kagome_metals_SC1,kagome_metals_SC2,kagome_metals_SC3}, for which the role of flat bands and correlation effects is investigated. 
Hence, there is an urgent need to further develop numerical methods to treat \gls{2D} systems. 
Important progress in describing strongly correlated \gls{2D} systems has been achieved by the development of methods, which help to investigate quantum phases and quantum critical behavior directly in the thermodynamic limit.
The fundamental insight is that the challenge of solving systems in the thermodynamic limit can be simplified by replacing it with an equivalent problem that is defined on a finite cluster, only.
This is the foundation of so\hyp called \glspl{CET} such as \gls{DMFT} \cite{metzner-vollhardt,Georges1992,Georges1996}, \gls{CPT} \cite{GrosValenti93, Senechal2000} or \gls{VCA} \cite{VCA1,VCA2,VCA3,VCA_ana}.
These approaches are often formulated in terms of Green's functions of the finite cluster and the embedding is defined such that the cluster Green's function describes best that of the actual physical system.
Despite others, the success of these methods is based on the intimate relation between Green's functions and experimental observables, for instance spectral functions, which allows for the direct comparison between theory and experiment \cite{hufner,RevModPhys.75.473}.
With the development of \glspl{CET}, which contain the mapping of lattice problems to finite clusters, the focus has shifted to determine the cluster Green's function, which still constitutes a computational highly non\hyp trivial problem.
Important classes of methods to approach this task are wave function\hyp based methods, like \gls{ED} \cite{noack:93NO,SandvikReview} or \gls{TNS} methods \cite{review_tensorproductstates_CiracVerstraete,MPSreviewVerstraete}, and sampling\hyp based methods, like \gls{QMC} \cite{QMC_review1,QMC_review2,Albuquerque20071187NO}.
\Gls{ED} cluster solvers are conceptionally the most simple approach and despite the fact that an extreme degree of optimization has been achieved, incorporating various symmetries and exploiting massive parallelization \cite{Wietek_Laeuchli2018}, they suffer drastically from the exponential growth of the Hilbert space, limiting the practically doable number of cluster sites for single\hyp band Hubbard models to $\sim \mathcal{O}(20)$.
On the other hand, \gls{TNS} realized as \gls{MPS} \cite{Schollwoeck201196} are extremely successful in one dimension, but the area law of entanglement or the entanglement barrier in non\hyp equilibrium situations highly limits their applicability to \gls{2D} systems \cite{Paeckel2019}.
From a quantum information perspective, finite \gls{PEPS} \cite{verstraete-cirac_2D,MPSreviewVerstraete} should be optimal.
However, these algorithms have been shown to exhibit only extremly slow convergence for Hubbard\hyp type systems \cite{Scheb}.
Finally, \gls{QMC} methods have their limitations due to the fermionic sign problem \cite{QMC_signproblem_Troyer}, which restricts the class of solvable cluster Hamiltonians.
In this paper, we introduce an efficient \gls{MPS}\hyp based scheme for calculating the cluster Green's function and apply it within the context of \gls{VCA}.
It turns out that a na\"ive treatment of established Krylov\hyp based methods in a \gls{MPS} framework would barely improve the range of applicability of the \gls{VCA}.
However, when combining the \gls{MPS} with a band Lanczos scheme as well as an energy truncation, it is possible to treat cluster geometries well outside the reach of \gls{ED} solvers and without principal limitations on the Hamiltonian.
The remainder of the paper is structured as follows: In \cref{sec:model} we introduce the single\hyp band Hubbard model and the basic idea of cluster methods, such as the \gls{VCA}.
In \cref{sec:method} we discuss in detail our band Lanczos ansatz for \gls{MPS}, including the energy truncation scheme, an error estimate using the Hochbruck\hyp Lubich bound \cite{lubich_timeevolve}, a useful energy rescaling, and a discussion of the loss of orthogonality of the Lanczos vectors within the \gls{MPS} framework.
Furthermore, \cref{sec:VCA} discusses further aspects of the \gls{VCA} as relevant for this paper, and \cref{sec:clusters} discusses our choice of clusters, including Betts clusters.
\Cref{sec:results} presents the main results of this paper obtained with our band Lanczos \gls{MPS}+\gls{VCA} ansatz, including a detailed analysis of the scaling of the results with the \gls{MPS} bond dimension and a comparison of the staggered magnetization after scaling to infinite cluster size with numerically exact \gls{QMC} results, as well as results for the spectral function.
We finish the results section by applying the band Lanczos \gls{MPS}+\gls{VCA} ansatz to calculate spectral properties of a more complex model as relevant to electron\hyp doped CaCuO$_2$.
In \cref{sec:conclusion} we conclude and provide an outlook to further applications of our method.
The appendix presents technical details of the calculations.
\section{\label{sec:model}Model and numerical technique}
The model for which we benchmark our solver is one of the most common ones to study strongly correlated electron systems in two dimensions, the one\hyp band Hubbard model with nearest\hyp neighbor hopping \cite{Hubbard1963}.
The corresponding Hamiltonian reads %
\begin{align}%
	\label{eq:OneBandHubbardModel}%
	\hat H = %
		-t \sum_{\braket{i,j}, \sigma} \left( \hat c^{\dagger}_{i, \sigma} \hat c^{\nodagger}_{j, \sigma} + \mathrm{h.c.} \right) + U \sum_{i} \hat n^{\nodagger}_{i, \uparrow} \hat n^{\nodagger}_{i, \downarrow}\; ,%
\end{align}%
where $\hat c^{(\dagger)}_{j, \sigma}$ are the annihilation (creation) operators of electrons of spin $\sigma$ and $\hat n^{\nodagger}_{j, \sigma} = \hat c^{\nodagger}_{j, \sigma} \hat c^{\dagger}_{j, \sigma}$ is the particle number operator for a given spin $\sigma$ on site $j$.
For our band Lanczos solver we have quantum cluster techniques in mind, for which a finite\hyp size scaling in cluster size is known to be notoriously difficult in \gls{2D} \cite{PhysRevX.5.041041,2DHubbard_Science}.
In particular embedding the cluster into an effective environment, various factors need to be considered, such as the ratio between cluster\hyp bulk and cluster\hyp boundary sites, the connectivity of boundary sites or the cluster geometry.
Given these complications, which we discuss in more detail in \cref{sec:clusters}, the main goal of our band Lanczos solver is to provide a tool that can handle various cluster geometries while increasing the cluster sizes as much as possible.
A prime example for a quantum cluster technique is \gls{CPT} \cite{GrosValenti93, Senechal2000}, which can be derived from strong\hyp coupling perturbation theory \cite{Pairault98,Senechal2002}, and which allows to calculate the spectral function based on the cluster Green's function.
In CPT, the lattice is tiled into a superlattice of -- most of the time identical -- clusters.
The inter\hyp cluster hopping terms are then collected into a matrix $\mathbf{V}$ and treated in lowest order perturbation theory when constructing the \gls{CPT} Green's function:
\begin{align}%
	\label{eq:CPT}
	\mathbf{G}^{\mathrm{CPT}}(\mathbf{\tilde{k}},\omega) = %
	\left( (\mathbf{G}^{\mathrm{cluster}}(\omega))^{-1} - \mathbf{V}(\mathbf{\tilde{k}}) \right)^{-1} .%
\end{align}%
Here $\mathbf{\tilde{k}}$ are reduced wavevectors corresponding to the partial Fourier transform with respect to the superlattice vectors.
In the following, we will omit the superscript and refer to the cluster Green's function as $\mathbf{G}(\omega)$.
The error of the perturbative treatment within \gls{CPT} is expected to reduce when enlarging the ratio of cluster 'bulk' to cluster boundary, for instance when scaling up clusters of fixed geometry.
In addition to the cluster size, different geometries allow to precisely account for the spectral function at different wave vectors $\mathbf{k}$.
\Gls{CPT} has been used not only for model calculations on the Hubbard \cite{Senechal2000}, $tJ$ \cite{Zacher2000} and pure spin models \cite{Ovchinnikov2010}, but also for more realistic modeling of materials, e.g., in the context of Mott insulators such as NiO \cite{Eder2005} or (doped) cuprates \cite{Dahnken2002,Senechal2002,Dahnken2003,Senechal2004}.
It can also be extended to non\hyp local interactions \cite{Aichhorn2004}, electron\hyp phonon coupling \cite{Hohenadler2003}, non\hyp equilibrium problems \cite{Balzer2011} or even be used to calculate two\hyp particle responses \cite{Kuzmin2023}.
\Gls{CPT} is thereby not only conceptually simple, but also one of the most versatile quantum cluster techniques.
However, \gls{CPT} is limited in that it does not allow to study symmetry\hyp broken phases.
An extension of the technique mends this deficiency by including symmetry\hyp breaking Weiss fields on the cluster, which are determined according to a variational principle.
The variational \gls{CPT}, more commonly known as \gls{VCA} \cite{VCA1,VCA2}, will be introduced in more detail in \cref{sec:VCA}.
For benchmarking purposes, we add here a Weiss field term on the cluster corresponding to a staggered magnetic field, %
\begin{align}%
	\label{eq:WeissField}%
	\hat{\mathcal{H}}_{\mathrm{AF}} = %
	h_z \sum_l e^{i\mathbf{Q}\cdot\mathbf{R}_l} %
	(\hat{n}_{\mathbf{R}_{l,\uparrow}}-\hat{n}_{\mathbf{R}_{l,\downarrow}}),%
\end{align}%
where $l$ denotes cluster sites and the N\'eel ordering wave vector is $\mathbf{Q}=(\pi,\pi)$.
A detailed \gls{VCA} study of \gls{AFM} in the \gls{2D} Hubbard model using this Weiss field can be found in Ref.~\onlinecite{VCA2}.
For the following, it is important to note that the (cluster) Green's function of $\hat{H}+\hat{\mathcal{H}}_{\mathrm{AF}}$ needs to be determined.
\section{\label{sec:method}Method}

In this section we introduce the band Lanczos algorithm for calculating the cluster Green's function using \gls{MPS}.
Several routes have been pursued so far to employ \gls{MPS}\hyp based cluster solvers for \gls{CET}.
This includes time\hyp evolution based solvers obtaining the cluster Green's function from real or in imaginary time evolution \cite{Linden2014,Ganahl2015,Bauernfeind2017,Bramberger2023,PhysRevX.13.011039}, expanding the resolvent in terms of orthogonal polynomials \cite{Holzner2011,Tiegel2014,Wolf2014}, reformulating the determination of the resolvent as optimization problem using correction vectors \cite{Kuehner1999,Nocera2022}, or evaluating the Green's matrix elements explicitly by means of a global Lanzcos approach \cite{Dargel2012}.
Nevertheless, despite each of these methods being considerably successful in the past, their applicability to \gls{VCA} is rather limited.
Standard time evolution based solvers require considerable system sizes to avoid too fast entanglement growth, due to reflections of excitations at the cluster boundary.
Evaluating the resolvent in terms of orthogonal polynomials on the other hand, typically requires a large number $\sim \mathcal O(100)$ of basis states, each of which is obtained by applying the Hamiltonian to its predecessor, and thereby rapidly increases the state's entanglement.
This limitation is partially overcome by a direct Lanczos expansion of each Green's matrix element, which typically requires only $\sim \mathcal O(10)$ applications of the Hamiltonian, per matrix element.
However, a lot of information about the low\hyp energy states is generated repeatedly, since each matrix element is constructed independently of every other.
Furthermore, the reduced numerical precision using \gls{MPS} arithmetics typically causes severe orthogonality problems after more than $10$ applications of the Hamiltonian.
The necessary application of further reorthogonalization techniques common to global Krylov methods \cite{Simon1984} deteriorates the computational efficiency such that cluster sizes beyond those doable using exact methods can only hardly be reached.
In sight of these complications, the band Lanczos \cite{Freund2000,Baker2023} offers a promising approach to evaluate the cluster Green's function in a global Krylov subspace representation, employing various initial states.
In the following, we first summarize the general idea of the band Lanczos and introduce the necessary steps to obtain a meaningful convergence criterium.
Afterwards, we recapitulate the necessary theory of the \gls{VCA}, in order to use the band Lanczos algorithm as its cluster solver.
\subsection{Band Lanczos with \gls{MPS}}
The quantity of interest of most \glspl{CET} is the cluster Green's function in frequency space, which can be written componentwise for electrons in real space as
\begin{align}
	G^{(e)}_{\mu\mu^\prime}(\omega)
	=
	\braket{\psi| \hat c^\nodagger_\mu \left(\hat H - E_0 - \omega \right)^{-1} \hat c^\dagger_{\mu^\prime} |\psi} \; , \label{eq:greens-function}
\end{align}
where we combined site and spin labels into greek letters $\mu = (i,\sigma), \mu^\prime=(j,\sigma^\prime)$ and the superscript $(e)$ indicates that here we are describing the electron part.
The cluster Green's function for holes, indicated by a superscript $(h)$, can be written in a similar manner by exchanging the creation and annihilation operators and replacing $\omega \rightarrow -\omega$.
In the band Lanczos method, the Green's function is calculated by constructing a Krylov subspace using several electron and hole excitations, which serve as initial states.
Representing the initial states as \gls{MPS} and the Hamiltonian as \gls{MPO}, the Krylov subspace construction can be performed in principle using standard \gls{MPS} arithmetics.
The particular choice of the set of initial states can depend on the cluster symmetries, but for the sake of simplicity we restrict ourselves in the following to the simplest case and ignore symmetries.
Consider the ground state $\ket{\psi}$ of a cluster, given by the Hamiltonian $\hat H$ acting on $L\in \mathbb N$ cluster sites, where each site $j$ can host electronic degrees of freedom described by annihilation (creation) operators $\hat c^{(\dagger)}_{j, \sigma}$.
We define a set of initial states $\ket{\varphi^{(e)}_{j,\sigma}} = \hat c^\dagger_{j, \sigma} \ket{\psi}$ and introduce the spin $\sigma$ electron Krylov subspace $\mathcal K^D_\sigma$ with $D=N \cdot L$, generated from $N-1$ applications of $\hat H$ to the $L$ initial states with spin $\sigma$:
\begin{align}
	\mathcal K^0_\sigma &= \operatorname{span} \left\{\ket{\varphi^{(e)}_{1,\sigma}}, \ldots, \ket{\varphi^{(e)}_{L,\sigma}} \right\} \; , \\
	\mathcal K^D_\sigma &= \mathcal K^{D-1}_\sigma \cup \operatorname{span} \left\{\hat H^N \ket{\varphi^{(e)}_{1,\sigma}}, \ldots, \hat H^N \ket{\varphi^{(e)}_{L,\sigma}} \right\} \; . 
\end{align}
Note that replacing the creation operators with the annihilation operators $\hat c^\dagger_{j, \sigma} \rightarrow \hat c^\nodagger_{j, \sigma}$, we obtain another set of initial states $\ket{\varphi^{(h)}_{j,\sigma}} = \hat c^\nodagger_{j, \sigma} \ket{\psi}$, from which we construct the spin $\sigma$ hole Krylov subspace $\bar{\mathcal K}^D_\sigma$.
For each Krylov subspace, an orthonormal basis is obtained from an iterative orthogonalization scheme.
Thereby, given a set of orthogonal Krylov vectors $\ket{k_{\alpha,\nu}}\in \mathcal K^D_\sigma$ with $\alpha\in\left\{1,\ldots,\tilde L \right\}$ ($\tilde L \leq L$) and $\nu \in \left\{0,\ldots,N-1 \right\}$, a new set of candidate states $\left\{\ket{k_{\alpha, N}}\right\}$ is generated by applying the Hamiltonian to all states $\left\{\ket{k_{\alpha, N-1}}\right\}$. 
Just as in the conventional Lanczos scheme \cite{CullumWilloughby}, every candidate state $\ket{k_{\alpha, N}}$ is then reorthogonalized against its $2L$ predecessors $\ket{k_{\alpha, N-1}}$ and $\ket{k_{\alpha, N-2}}$.
Furthermore, candidate states have to be orthogonalized against each other, i.e., upon adding a new candidate state $\ket{k_{\alpha, N}}$, it needs to be orthogonalized against all $\ket{k_{\alpha^\prime < \alpha, N}}$.
It may occur that the states $\ket{k_{\alpha, N}}$ generated by this recursion are not linearly independent.
This situation is typically solved by applying a so\hyp called deflation scheme, reducing the range $\tilde L$ of the $\alpha$'s until all Krylov states $\ket{k_{\alpha\leq \tilde L, N}}$ are linearly independent \cite{Freund2000}.
Following this strategy, a global Krylov basis is constructed that allows for a representation of the Green's function \cref{eq:greens-function} in a Krylov subspace
\begin{align}
	G^{(e)}_{\mu\mu^\prime}(\omega)
	&\approx
	\sum_{\nu,\alpha_\nu} \sum_{\nu^\prime,\alpha^\prime_{\nu^\prime}}
	\braket{\psi|\hat c^\nodagger_\mu|k_{\alpha^\noprime_\nu,\nu}} \braket{k_{\alpha^\prime_{\nu^\prime},\nu^\prime}|\hat c^\dagger_{\mu^\prime}|\psi} \notag \\
	&\phantom{= \sum_{\nu,\alpha_\nu}}
	\times \braket{k_{\alpha^\noprime_\nu,\nu}|\left( \hat H - E_0 - \omega \right)^{-1}|k_{\alpha^\prime_{\nu^\prime},\nu^\prime}} \; . \label{eq:greens-function:krylov}
\end{align}
In absence of deflated states, the effective Hamiltonian $\braket{k_{\alpha, \nu} | \hat H | k_{\alpha^\prime, \nu^\prime}}$ is block tri\hyp diagonal with a maximum block size $2L+1$ and can be diagonalized easily, to evaluate the operator inverse.
In contrast, the usual Lanczos recursion corresponds to the case $L=1$:
The block size is $2\cdot 1 + 1 = 3$, which reflects the fact that only a single matrix element of the cluster Green's function can be evaluated per Krylov expansion.
The overall dimension of the generated Krylov subspace is given by $N\cdot L$, which amounts to $N-1$ applications of the Hamiltonian, and we refer to $N$ as the Krylov order.
This, however, comes at the cost of additional, global \gls{MPS} arithmetics when reorthogonalizing the candidate states.
The resulting approximations generally introduce instabilities in the band Lanczos recursion.
Instabilities most prominently manifest in form of a loss of the block tri\hyp diagonal structure of the effective Hamiltonian.
In other words, weight of the basis states accumulates in the orthogonal complement of the targeted Krylov subspace.
The loss of accuracy of the Krylov subspace expansion leads in particular to two problematic issues.
First, the loss of orthogonality amongst the basis states in the Krylov subspace cannot be completely compensated in practice:
Due to the large number of basis states, a (full) reorthogonalization would not be efficient numerically.
Second, a convergence criterion that relates error bounds of the cluster Green's function to the approximation quality of the Krylov subspace is required: Standard convergence criteria such as the relative change in the energy spectrum are numerically unstable.
Our methodical developments target both convergence issues with the goal of a numerically stable band Lanczos recursion using \gls{MPS} arithmetics that work at a finite, yet controlled approximation quality.
\subsubsection{Energy truncation}
\begin{figure}[!h]
	\centering
	\subfloat[\label{fig:e-trunc:sketch}]{
		\includegraphics[height=0.25\textheight]{\imgpath/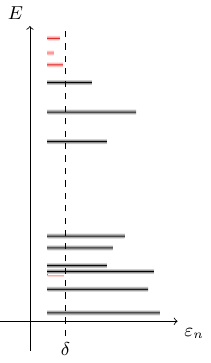}
	}%
	\subfloat[\label{fig:e-trunc:dos}]{
		\includegraphics[height=0.25\textheight]{\imgpath/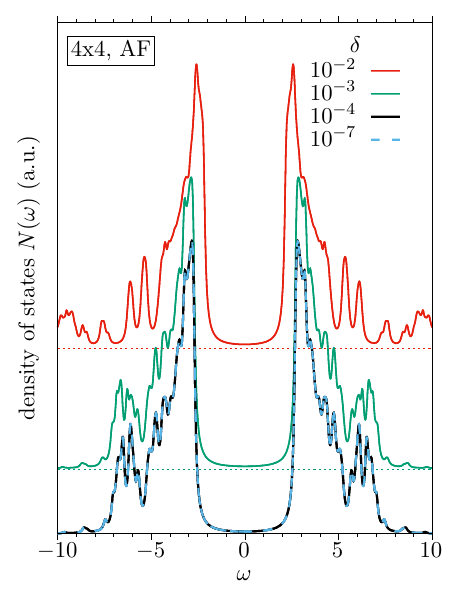}
	}
	\caption
	{
		\label{fig:Etrunc}
		\subfigref{fig:e-trunc:sketch} Schematic representation of the energy truncation.
		Overlaps $\epsilon_n=\langle\tilde{E}_n\vert\psi\rangle<\delta$ are neglected when constructing the sublattice projection (marked red). 
		\subfigref{fig:e-trunc:dos} Effect of the truncation threshold on the (lattice) density of states $N(\omega)$ for a $4\times4$ cluster in the \gls{AFM} phase, see \cref{sec:VCA} for details.
		The results for $\delta=10^{-2},10^{-3}$ are shifted for clarity.
	}
\end{figure}
Constructing Krylov spaces of higher order, an increasing number of highly excited states has to be captured by the \gls{MPS} representations of the Krylov vectors.
Unfortunately, states from the bulk of the many\hyp body Hamiltonian's spectrum typically satisfy a volume law of entanglement \cite{Bianchi2022} and are therefore the main reason for the exponentially increasing computational complexity.
While this is a problem in general, here, we are interested in a faithful approximation of the poles of the Green's matrix, i.e., their positions and weights.
Hence, we can exploit a strategy introduced previously in the context of expanding the resolvent in terms of Chebyshev polynomials \cite{Holzner2011}.
The general idea is to construct local Krylov spaces at each lattice site and iteratively project the state's site tensors into a sequence of local Krylov subspaces.
However, in contrast to projecting the site tensors into a certain energy window as described in \cite{Holzner2011}, we use the local Krylov space expansion to construct a projector to the subspace containing the most relevant eigenstates, up to a defined threshold $\delta$:
\begin{align}
	\ket{\psi}
		=
			\sum_{n}\underbrace{\braket{\tilde E_n|\psi}}_{\varepsilon_n} \ket{\tilde E_n}
			\approx 
			\sum_{\varepsilon_n > \delta} \varepsilon_n \ket{\tilde E_n}
\end{align}
where
\begin{align}
	\ket{\tilde E_n} = \sum_{l=0}^{K} c_l \ket{\nu_l} \; , \quad \lim_{K\rightarrow \operatorname{dim} \mathcal H} \hat H \ket{\tilde E_n} = E_n \ket{E_n}
\end{align}
and $\ket{\nu_l}\in \mathcal H_K$ Lanczos vectors in Krylov space $\mathcal H_K = \operatorname{span} \left\{\ket{\psi}, \hat H \ket{\psi} , \ldots, \hat H^K \ket{\psi}\right\}$.
A schematic illustration of the energy truncation is shown in \cref{fig:e-trunc:sketch}.
The red shaded bars correspond to those eigenstates that have a weight $\braket{\tilde E_n|\psi} < \delta$ and are hence discarded.
Refering to the expansion of the Green's function in the Krylov subspace \cref{eq:greens-function:krylov}, we can identify the weights with the expansion of the matrix elements $\braket{k_{\alpha^\prime_{\nu^\prime},\nu^\prime}|\hat c^\dagger_\mu|\psi}$ in the energy eigenbasis.
The energy truncation thus simply discards those eigenstates, whose contribution to the total Green's function are $\mathcal O (\delta^2)$.
In \cref{fig:e-trunc:dos} we illustrate the effect of the energy truncation on the calculated lattice \gls{DOS} using the \gls{CPT} Green's function for a $4\times4$ cluster in the antiferromagnetically ordered state.
Setting the truncation threshold too high results in a breakdown of the subspace projection.
Here, this is the case for $\delta=10^{-2}$, where the erroneous subspace projection leads to a \gls{DOS}, which neither captures correctly the gap nor the high\hyp frequency part of the spectrum.
However, already for moderate values of $\delta\lesssim10^{-3}$ the spectrum is well reproduced, in particular the gap and the low\hyp energy part of the spectrum, and only small deviations from the exact result are visible for higher excitation energies.
A truncation threshold of $\delta\leq10^{-4}$ finally leads to results which are for practical purposes essentially exact.
In this case, only states with negligible weight for the density of states are truncated.
The systematic improvements of the high\hyp energy part of the spectrum can be related to the fact that the Green's function is computed from locally exciting the ground state, such that the weight of higher-energy eigenstates decreases rapidly.
For the following calculations, we avoid any ambiguity by using an energy truncation threshold of $\delta=10^{-5}$.
The computational costs of this energy truncation are not negligible.
The complexity per local update scales as $\mathcal O(K m^3 w d) + \mathcal O(K m^2 w^2 d^2)$ where $K$ is the number of applications of the local representation of $\hat H$.
Refering to standard notation, here we denote the \gls{MPS} bond dimension by $m$, the \gls{MPO} bond dimension by $w$ and the local Hilbert space dimension by $d$.
Thus, we use the energy truncation only for the large clusters with $L\geq16$ sites and apply it in the process of creating a new candidate state only once, namely after the global application of $\hat H$.
However, the goal is not to achieve a speed up at constant $m$.
Instead, we aim for a reduction of orthogonality losses that occur when orthogonalizing candidate states with respect to each other.
Here, the effect of the energy truncation is to remove highly entangled contributions allowing for signficantly higher precision in the orthogonalization procedure, while keeping the bond dimension fixed.
We found that building up a reasonably large Krylov space of dimension $D\sim 100$ for the larger clusters using computationally feasible bond dimensions $m=1024-2048$ was possible only when employing an energy truncation.
Otherwise, orthogonality could not be maintained and the construction procedure would break down due to orthogonality losses.
\subsubsection{\label{sec:method:bl:hl}Hochbruck\hyp Lubich criterion}
\begin{figure}[t]
	\centering
	\includegraphics[width=\linewidth]{\imgpath/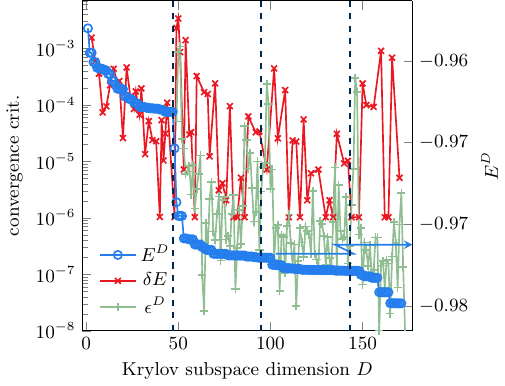}
	\caption{
		\label{fig:convergence-criteria}
		Comparison of convergence critera.
		The blue curve (right $y$\hyp axis) shows the lowest eigenvalue $E^D$ of the effective Hamiltonian in the Krylov subspace as a function of the dimension $D$ for a $4\times 6$\hyp site Hubbard cluster at $U=8t$ and a staggered magnetic field $h_z=0.09t$.
		The conventional convergence criterion $\delta E = E^{D+1} - E^{D}$ is given by the red line (left $y$\hyp axis), exhibiting strong fluctuations over several orders of magnitude.
		In contrast, oscillations in the suggested Hochbruck\hyp Lubich critertion $\epsilon^D$ shown by the green line are significantly smaller, overlaying a systematic decrease with respect to the number of iterations.
	}
\end{figure}
Controlling the convergence of the band Lanczos in the \gls{MPS} representation is a suprisingly delicate problem.
The unavoidable loss of numerical precision, caused by performing global operations using \gls{MPS} arithmetics, as well as finite truncation errors, render standard convergence criteria such as the relative change in the ground state energy in the constructed Krylov subspace not suitable.
For instance, initially one observes a steep decrease of the relative error $\delta E = (E^{D+\alpha+1} - E^{D+\alpha})/E^{D+\alpha}$, where $E^{D+\alpha}$ is the ground state energy after adding the $\alpha$th candidate state to the Krylov subspace $\mathcal K^D_\sigma$.
However, $\delta E$ exhibits rather drastic jumps over more than an order of magnitude, paired with iterations where $\delta E=0$ with respect to the energy resolution, which is bounded by the \gls{MPS} truncation error.
This phenomenon is shown at the example of a $4\times 6$\hyp Hubbard cluster in \cref{fig:convergence-criteria} where the red curves represent $\delta E$.
There are two major sources for an erroneous convergence indication in $\delta E$.
First, whenever $\hat H$ is applied to a complete sequence of candidate states (here, $N \cdot 2L = N\cdot 48$ marked by the dashed lines), the Krylov order $N$ increases.
Given a fixed Krylov order, the band Lanczos generates a sequence of states that optimize the approximation of the ground state, which, however, is constraint by the Krylov order.
Therefore, $\delta E$ decreases within a sequence of candidate states, which leads to a false indication of overall convergence.
Second, candidate states $\ket{k_{\alpha, N}}$ have a different relevance for the ground state approximation.
Most prominently, states that are created from initial excitations located at the boundary of the system converge more quickly.
Besides these technical problems, in practise the energy gain is not the most relevant quantity: In the end, the goal is to approximate the poles and weights of the cluster Green's function up to a certain precision.
For that purpose, here we pursue a different approach using a convergence criterion that directly measures the desired approximation quality.
The idea is to relate an error measure $\delta R^D(\tau)$, which can be derived for the approximation of the action of the operator exponential $\mathrm e^{-\mathrm i \hat H \tau} \ket{\psi}$ to the Green's function.
Here, $\delta R^D(\tau)$ denotes the generalized residual introduced by Hochbruck and Lubich \cite{lubich_errors,lubich_timeevolve,Botchev2013}, which bounds the actual approximation error of the action of the operator exponential via $\delta R^D(\tau) \leq \epsilon^D$, and an estimation for $\epsilon^D$ is given by the Hochbruck\hyp Lubich bound \cite{lubich_timeevolve}
\begin{align}
	\epsilon^D_\tau \le \frac{40}{\tau} e^{-\frac{\tau}{4}}\left(\frac{e \tau}{4D}\right)^D\;. \label{eq:hl}
\end{align}
Note the exponential suppression of the bound on the time step $\tau$.
We propose to use the generalized residual for two reasons.
(1) Evolving for small time steps $\tau$, the approximation of the time evolution incorporates mostly low\hyp lying excited states because the spread of information is limited by the Lieb\hyp Robinson bounds~\cite{Nachtergaele2006} and $\ket{\psi}$ is close to the ground state.
(2) $\delta R^D(\tau)$ is remarkably insensitive to small approximation errors in the low-energy spectrum, in particular if poles move close to each other in the scope of the Krylov sequence.
In our computations, we choose $\epsilon \leq \delta R^D(\tau)$ as error threshold, which can be evaluated at negligible numerical costs from the matrix representation of the exponential of the effective Hamiltonian $\hat H^D$ \cite{Botchev2013,Paeckel2019} (see also \cref{app:hl}).
We furthermore fix the time step by relating it to the spectral width $W$ of $\hat H$ via $\tau = \frac{2\pi}{\kappa W}$ where typically we choose the \textit{energy window fraction} $\kappa \approx 1$.
We can now argue why we expect this error measure to be more reliable than only considering the relative change in the groundstate energy:
$\delta R^D(\tau)$ is not a relative measure and while we also found it to exhibit fluctuations in our numerics, these are less severe than those in $\delta E$.
The stability of $\delta R^D(\tau)$ against fluctuations comes from the fact that increasing the Krylov space dimension $D$, the residual in approximating the time evolution is monotonically decreasing, which follows directly from \cref{eq:hl}.
Thus, fluctuations can only be generated from approximation errors caused by the \gls{MPS} arithmetics and truncation errors.
The main effect of these errors is to destroy the block\hyp tridiagonal form of the effective Hamiltonian, which, however, can be monitored by the precision of the \gls{MPS} arithmetics and the truncation errors.
We choose the corresponding thresholds to be sufficiently small and thereby obtain a significantly more stable convergence criterion than using $\delta E$.
It should be noted that strictly speaking,~\cref{eq:hl} is valid only after a complete sequence of Lanczos states for the set of initial states $\ket{\varphi^{(e)}_{j,\sigma}}$ has been constructed, i.e., the Krylov space dimension has to fulfill $D=N\cdot L$.
However, from our numerical experience we found that the violation of monotonicity while constructing a new set of $L$ candidate states is not too severe.
\subsubsection{Orthogonality loss}
A major source for numerical instabilities in Krylov subspace methods is the loss of orthogonality of Krylov vectors.
While in the commonly used exact state representations the loss of orthogonality is in general caused by round\hyp off errors due to finite precision arithmetics \cite{Meurant2006}, the use of \gls{MPS} introduces additional error sources.
On the one hand, generating a new set of candidate states in the $N$th iteration of the band Lanczos recursion requires to act with the Hamiltonian on \gls{MPS} representations of Lanczos vectors $\ket{k_{\alpha, N}}$: $\ket{k_{\alpha, N}} \longmapsto \hat H \ket{k_{\alpha, N}}$.
Performing this operation by a na\"ive \gls{MPO}\hyp \gls{MPS} application is numerically very costly so that we resort to a variational application scheme with a zipup preconditioner \cite{Paeckel2019}.
While this allows us to achieve convergence of the \gls{MPO} application typically after two sweeps, we are still limited by the growth of the \gls{MPS} bond dimension, requiring a finite truncated weight.
On the other hand, the orthogonalization of new candidate vectors $\hat H \ket{k_{\alpha, N}}$ against previous Lanczos states has to be done using a variational update scheme at finite truncated weight, too.
This introduces a further loss of numerical precision if the variational optimization can get stuck in local minima.
We preconditioned the optimization using an optimized initial guess state generation by mixing small contribution of higher and lower order Krylov states into the candidates.
Nevertheless, the band Lanczos recursion inevitably is going to loose orthogonality, and the rate at which this loss occurs crucially depends on the maximally allowed truncated weight.
In order to keep the recursions stable as long as possible, we exploit the energy truncation, which removes poles, i.e., eigenstates, with small weights in the Green's function from the Lanczos states.
This truncation then allows us to use small truncated weights $\delta_\mathrm{trunc} \sim \mathcal{O}(10^{-10})$ at moderate bond dimensions.
We typically allow for up to $\chi_\mathrm{max}=1024$ states and only for large clusters increased that value, typically to $\chi_\mathrm{max}=2048$ and in case of the $6\times6$ cluster shown in \cref{Sec:FSscaling} up to $\chi_\mathrm{max}=2560$.
Throughout the band Lanczos recursion, we then monitor the violation of the anticommutation relations
\begin{align}
	\mathcal I_{j,\sigma;j^\prime,\sigma^\prime}
	&=
	\sum_{\alpha,\nu}
	\braket{0|\hat c^\dagger_{j,\sigma}|k_{\alpha,\nu}} \braket{k_{\alpha,\nu}| \hat c^\nodagger_{j^\prime,\sigma^\prime}|0} \notag \\
	&\phantom{=}\quad
	+
	\braket{0|\hat c^\nodagger_{j^\prime,\sigma^\prime}|k_{\alpha,\nu}} \braket{k_{\alpha,\nu}| \hat c^\dagger_{j,\sigma}|0} \; .
\end{align}
We consider a deviation
\begin{align}
	\delta_\mathrm{ol}
	=
	\sqrt{\sum_{j,j^\prime}\sum_{\sigma,\sigma^\prime} \lVert \mathcal I_{j,\sigma;j^\prime,\sigma^\prime} - \delta_{j,j^\prime}\delta_{\sigma,\sigma^\prime}\lVert^2} < 10^{-3} \; ,
\end{align}
which is of the order of the total truncation error $\sim L \cdot \sqrt{\delta_\mathrm{trunc}}$, as acceptable and terminate the Lanczos recursion if that threshold is exceeded.
For all practical applications discussed in this paper, the energy truncation allowed us to perform $\sim\mathcal{O}(100)$ Lanczos iterations without facing numerical instabilities caused by the orthogonality loss.
Importantly, this number was sufficient to converge also the largest clusters and obtain faithful approximations of the relevant poles and weights of the cluster Green's function with respect to the convergence criterion introduced in \cref{sec:method:bl:hl}.
\subsubsection{Energy rescaling}
In order to conveniently use the error bound \cref{eq:hl} we always rescale the Hamiltonian such that the spectral width is given by $W\equiv 1$.
Furthermore, the rescaling allows us to introduce another tool to control the quality of the constructed Krylov subspace.
In fact, a severe loss of orthogonality in the band Lanczos is typically signalled by an artificial drop of the lowest eigenvalue $\tilde E^D_{0}$ of the effective Hamiltonian in the Krylov space $\mathcal K^D$ below the groundstate energy.
For these reasons we rescale and shift the Hamiltonian $\hat H$ throughout our computations, which of course needs to be compensated for in the evaluation of the cluster Green's function.
The first step of the rescaling is to obtain the spectral width
\begin{align}
	W = \lvert E_0 - E_{\mathrm{max}} \rvert
\end{align}
of $\hat H$ by calculating the groundstate energy $E_0$ and the energy of the highest excited state $E_{\mathrm{max}}$.
Afterwards, the Hamiltonian is rescaled and shifted,
\begin{align}
	\hat H \rightarrow \hat{\tilde H} = \frac{\hat H - E_{\mathrm{max}}}{\lvert E_0 - E_{\mathrm{max}}\rvert} \;,
\end{align}
to force the spectrum to be located within the interval $[-1, 0]$.
Thus, the lowest eigenvalue of the effective Hamiltonian serves as additional proxy to the convergence of the Krylov space by monitoring whether the condition $\tilde E^D_{0} > -1$ holds true.
We also note that using the rescaled Hamiltonian, the Hochbruck\hyp Lubich time step is related trivially to the energy window fraction $\tau = 2\pi / \kappa$. 
\subsection{\label{sec:VCA}\gls{VCA}}
The \gls{VCA} \cite{VCA1} is an established quantum cluster technique, which is well\hyp suited to probe correlated systems for symmetry\hyp breaking fields \cite{VCA2,VCA3}.
It is based on the \gls{SEF} theory \cite{SEF1,SEF2,SEF3} and consists of determining the stationary points of the \gls{SEF} $\Omega_{\mathbf{t}}[\mathbf{\Sigma}]$, which is related to the Baym\hyp Kadanoff\hyp Luttinger\hyp Ward functional \cite{BaymKadanoff,LuttingerWard}, with respect to trial self\hyp energies of a reference system $\mathbf{\Sigma}(\mathbf{t}^{\prime})$:
\begin{align}
	\Omega_{\mathbf{t}}\left[\mathbf{\Sigma}\left(\mathbf{t}^{\prime}\right)\right] =& \Omega^{\prime} + \Tr \ln\left(-\left(\mathbf{G}_0^{-1}-\mathbf{\Sigma}\left(\mathbf{t}^{\prime}\right)\right)^{-1}\right) \notag \\
	&- \Tr\ln\left(-\mathbf{G}^{\prime}\right) \;.
	\label{eq:SEF}
\end{align}
Here, $\Tr \mathbf{A}=T\sum_{\omega,\alpha}A_{\alpha,\alpha}(i\omega)$ at temperature $T$, $\Omega^{\prime}$ denotes the grand potential, and $\mathbf{G}^{\prime}$ the Green's function of the reference system.
Since the theory requires the reference system to have the same interaction terms as the original system, the cluster self\hyp energies can be varied via their one\hyp body terms $\mathbf{t}^{\prime}$.
At a stationary point, the \gls{SEF} represents an approximation of the grand potential of the original system in the variational space of available self\hyp energies of the reference system.
To investigate symmetry\hyp broken phases, a (fictitious) symmetry\hyp breaking Weiss field term can be added to the cluster Hamiltonian. 
The Weiss field strength is then one of the one\hyp body terms, which are determined via the variational principle that leads to stationarity of the \gls{SEF} with respect to the cluster self\hyp energy.
Here, we focus on zero temperature, $T=0$, and perform the summation over frequency $\omega_m$ in \cref{eq:SEF} analytically.
In this case one obtains \cite{VCA_ana}
\begin{align}
	\Tr\ln(\mathbf{G}_0^{-1}-\mathbf{\Sigma})^{-1} &= \sum_m \omega_m\Theta(-\omega_m) - R\label{eq:sum1}\\
	\Tr \ln \mathbf{G^{\prime}}^{-1} &= \sum_m \omega_m^{\prime}\Theta(-\omega^{\prime}_m) - R \; ,\label{eq:sum2}
\end{align}
with the Heaviside function $\Theta(\omega)$, a contribution $R$ from the poles of the self\hyp energy, the poles $\omega^{\prime}_m$ of the cluster Green's function $\mathbf{G}^{\prime}$, and the poles $\omega_m$ of the \gls{VCA} Green's function $(\mathbf{G}_0^{-1}-\mathbf{\Sigma})^{-1}$.
It is therefore useful to work with a Lehmann representation of the Green's function since it explicitly includes the information about its poles $\omega_m$; for more details see \cref{app:Green}.
In order to limit the variational space of cluster self\hyp energies, a limited set of one\hyp body parameters of the cluster Hamiltonian are varied in practice: 
The hopping strengths $t^{\mathrm{cluster}}_{ij}$ can be determined from the variational principle to account for cluster\hyp boundary effects \cite{VCA1}, whereas the cluster chemical potential $\mu^{\prime}$ should be optimized to obtain a thermodynamically stable electron filling $n$ \cite{Aichhorn2006}.
Additional one\hyp body terms can be added to the cluster Hamiltonian to account for symmetry\hyp broken phases such as magnetic order \cite{VCA2}, superconductivity \cite{VCA3} or charge order \cite{Aichhorn2004}.
In the half\hyp filled \gls{2D} Hubbard model with isotropic nearest\hyp neighbor hopping, the chemical potential is at $\mu=U/2$ and $\mu^{\prime}$ does not need to be determined in the variational search.
Furthermore it was shown that optimizing the cluster hopping terms did not lead to significant improvement of the approximation of the grand potential \cite{VCA1}. 
To capture the essential physics of \gls{AFM} ordering, it is therefore sufficient to apply a staggered magnetic field of strength $h_z$ on the cluster, see \cref{eq:WeissField}, and use it as the sole one\hyp body parameter throughout the variational search.

\subsection{\label{sec:clusters}Choice of clusters}
\begin{figure}[tb]
	\includegraphics[width=\linewidth]{\imgpath/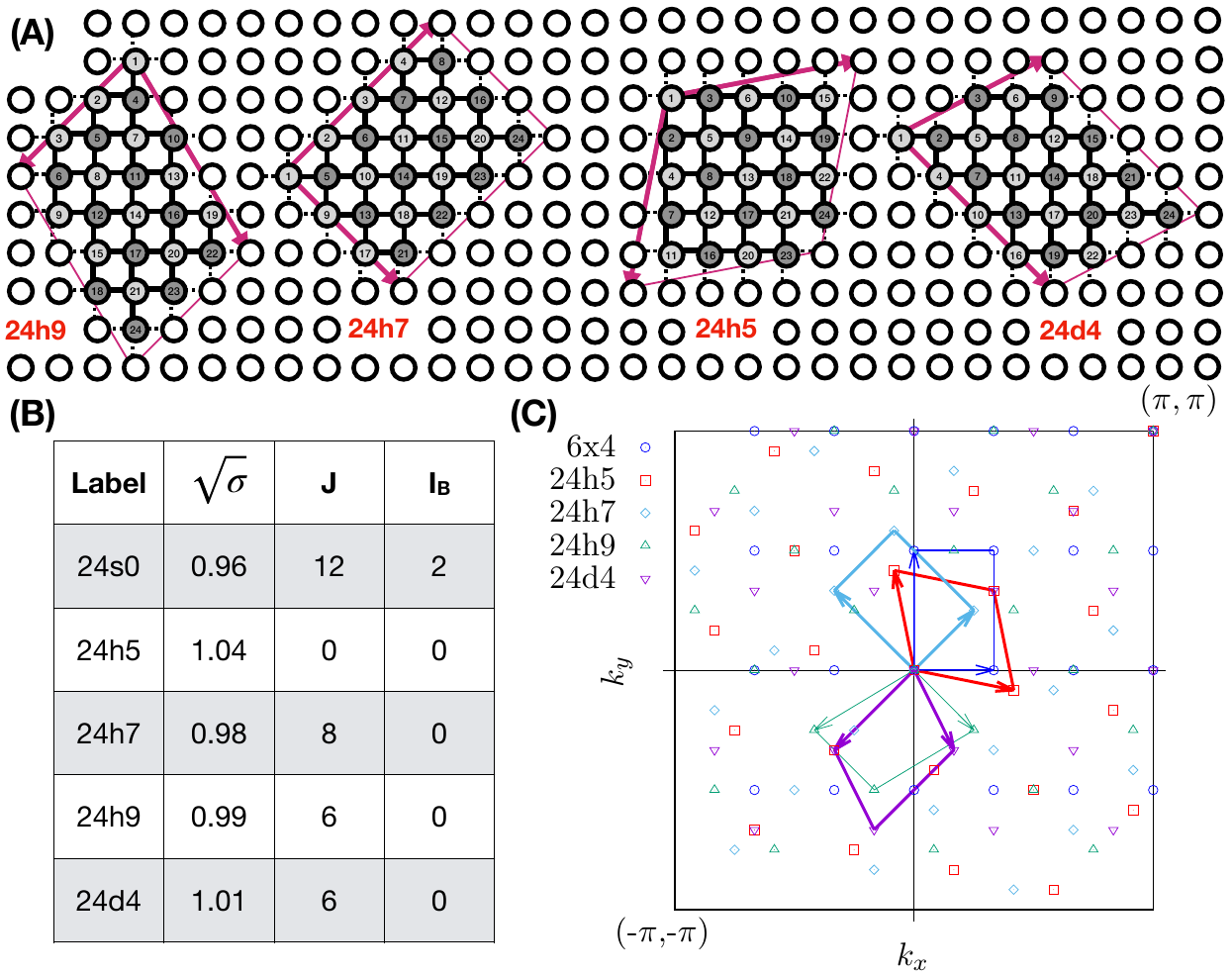}
	\caption
	{
		\label{fig:BettsSketch}
		(A) Illustration of the geometries of different clusters with $N=24$ sites labeled according to \textit{Betts et al.} \cite{Betts1996}.
		Red arrows indicate the superlattice vectors that tile the lattice with the respective cluster, dashed lines indicate inter\hyp cluster hopping terms and the N\'eel-type sublattice structure is indicated by different tones of grey.
		The clusters differ in their squareness ${\sigma}$, geometrical imperfection $J$ and bipartite imperfection $I_B$, see (B). 
		(C) Points in the Brillouin zone that are multiples of reciprocal superlattice vectors, which are indicated by arrows. 
		The corresponding reduced (superlattice) Brillouin zones are marked by parallelograms.
		An illustration of the corresponding \gls{1D} mapping is shown in \cref{app:1D}.
	}
\end{figure}

The choice of a cluster is a centerpiece of most quantum cluster techniques be it for accessing specific cluster momenta or for performing a finite\hyp size scaling.
Besides the size of the cluster, its geometry is of crucial importance in \gls{2D} systems and the seminal papers of Betts \textit{et al.} \cite{Betts1996,Betts99} introduced a set of evaluation criteria for the suitability of clusters with periodic boundary conditions, which were based on the completeness of near\hyp neighbor shells.
Within \gls{CPT} and \gls{VCA} one uses open boundary conditions on the clusters and a systematic study of Betts clusters is missing.
Nevertheless, we resort here to the labelling and characterization scheme introduced by \textit{Betts et al.} \cite{Betts1996}, in particular to parameters such as the geometrical imperfection $J$ and the bipartite imperfection $I_B$ to identify clusters of high geometrical and topological quality.
Apart from more traditional rectangular clusters, we also test our solver on skewed Betts clusters.
We limit ourselves to clusters that can tile bipartite lattices and focus on those which have zero bipartite imperfection.
By combining calculations of several cluster geometries, we can access different points of the lattice Brillouin zone via reciprocal vectors of the superlattice.
In \cref{fig:BettsSketch} we illustrate this for lattice tilings using different 24-site clusters that fulfil the condition $I_B=0$.
Apart from the accessible wave vectors $\mathbf{k}$, the choice of the tiling also affects the complexity of the \gls{MPS} calculation through the (optimized) mapping of the cluster sites onto a \gls{1D} chain, which we will discuss below.
\label{app:Betts}
\section{\label{sec:results}Results and discussion}
\begin{figure}[tb]
	\centering
	\includegraphics[width=\columnwidth]{\imgpath/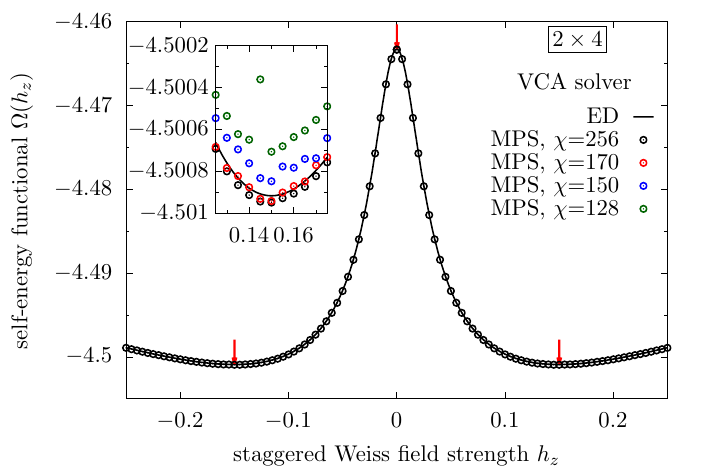}
	\caption
	{
		\label{fig:2x4}
		The \gls{SEF} $\Omega$ as a function of staggered Weiss field strength for a $2\times4$ cluster.
		Stationary points of $\Omega$ are indicated by arrows.
		Dots denote results obtained with the\gls{MPS}\hyp based solver, in the inset shown for different maximal \gls{MPS} bond dimension $\chi^{\mathrm{max}}$ around the minimum of $\Omega$; the results obtained with an \gls{ED} solver are shown as a line.
	}
\end{figure}

\begin{figure}[tb]
	\centering
	\includegraphics[width=\columnwidth]{\imgpath/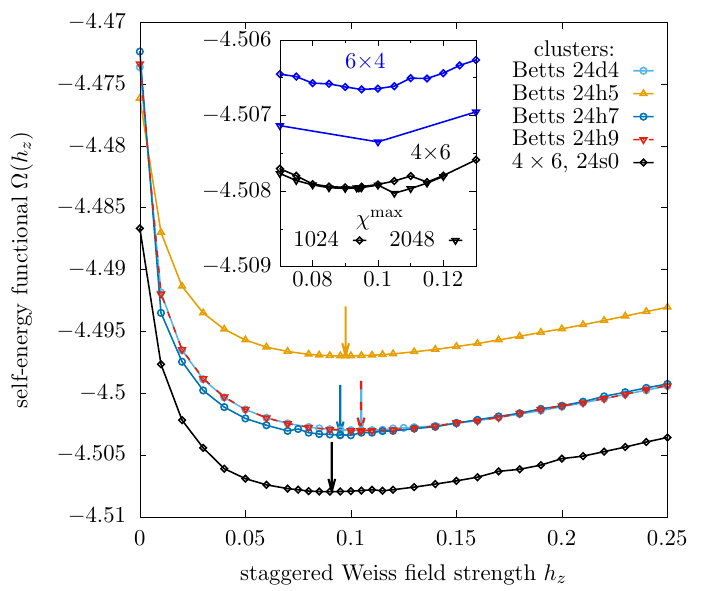}
	\caption{\label{fig:24clusters}
	Impact of the cluster geometry on the \gls{SEF}.
	$\Omega$ is calculated with $\chi^{\mathrm{max}}=1024$ for different 24-site clusters shown in \cref{fig:BettsSketch}.
	Minima of $\Omega$ are indicated by arrows, lines are only a guide to the eye.
	The inset shows $\Omega$ for two different \gls{1D} mappings of the 24s0 cluster, '$4\times6$' and '$6\times4$', calculated with $\chi^{\mathrm{max}}=1024, 2048$.
	}
\end{figure}

In the following, we study the \gls{AFM} ordering within the half\hyp filled single\hyp band Hubbard model to benchmark our solver. 
We focus here at the challenging point in parameter space $U/t=8$, i.e., an on-site interaction equal to the bandwidth of the non\hyp interacting dispersion, where the system is already in the Mott insulating regime. 
Translated to the \gls{SEF}, this leads to one stationary point, a maximum, at $h_z=0$ corresponding to a non\hyp magnetic solution and to two minima at $h_z=\pm h_z^c\neq0$ corresponding to a phase with \gls{AFM} order, see \cref{fig:2x4}.
Since $\Omega$ is symmetric in $h_z$, we plot it in the following only for positive values of $h_z$.
Compared to numerically exact solvers, the \gls{MPS}\hyp based solver introduces additional approximations, in particular since the \gls{MPS} bond dimension is bound to a maximal value $\chi^{\mathrm{max}}$ and since perfect orthogonality between the Krylov vectors is lost faster at large iteration numbers within the band Lanczos procedure.
Comparing the \gls{SEF} computed with our \gls{MPS}-based solver to the benchmark using an \gls{ED}\hyp based solver, we see excellent agreement in \cref{fig:2x4} if the maximal \gls{MPS} bond dimension is chosen large enough.
For the small cluster used here, $\chi$ can be increased until the \gls{MPS} representation is essentially exact.
However, for too small bond dimension, deviations from the exact curve are visible.
In that case, the \gls{SEF} is systematically evaluated to be too large.
Except for extreme cases, e.g. $\chi=128$ in \cref{fig:2x4}, $\Omega$ shows even for moderate bond dimension correct functional behavior as a function of the Weiss field, which allows for an interpolation around the stationary points.
As we show in \cref{sec:ChiScaling}, fitting the \gls{SEF} to determine its stationary points allows to calculate observables for different bond dimension in order to perform an extrapolation to $\chi^{\mathrm{max}}\rightarrow\infty$.
To illustrate the effects of the cluster geometry on $\Omega$, we plot in \cref{fig:24clusters} the \gls{SEF} of the 24-site clusters introduced in \cref{fig:BettsSketch}.
Even though the rough functional forms of $\Omega$ do not differ qualitatively, their stationary points vary notably.
We also note that clusters with similar geometric properties, as for instance quantified in terms of their geometrical and bipartite imperfection, see \cref{fig:BettsSketch}(B), generate quite similar \glspl{SEF}, whereas different imperfection of two clusters leads to significantly different $\Omega$.
Another consequence of using a \gls{MPS} representation of the quantum states is that even for a given cluster geometry, the obligatory mapping of the \gls{2D} cluster onto a \gls{1D} chain introduces a degree of freedom which has a direct impact on the precision with which the \gls{SEF} is calculated. 
In the inset of \cref{fig:24clusters} we illustrate this effect by calculating the cluster Green's function of the rectangular 24s0 cluster for two different 1D mappings. 
The corresponding \glspl{SEF} agree in the $\chi\rightarrow\infty$ limit, but differ for finite $\chi$.
Whereas the mapping minimizing long\hyp range hopping ('$4\times6$') leads to well converged results at  $\chi^{\mathrm{max}}=1024$, the longer\hyp ranged mapping ('$6\times4$') leads to a larger build\hyp up of entanglement and would require a $\chi^{\mathrm{max}}>2048$.
The range of the hopping terms after a mapping to a \gls{1D} chain and the corresponding bipartite entanglement entropy of the ground state are discussed in \cref{app:1D} for the different 24-site clusters shown in \cref{fig:BettsSketch}.

In presence of nearly degenerate quantum states, different symmetry\hyp breaking orders on the cluster can be easily picked up within \gls{MPS} based approaches when not providing a sufficiently large bond dimension.
This is for instance the case for the '$4\times6$' mapped 24s0 cluster at $h_z\approx0.11$, see inset of \cref{fig:24clusters}.
Hence, scaling the position of the stationary points of $\Omega$ in the maximal \gls{MPS} bond dimension instead of individual points of the \gls{SEF} proved to be more robust.
\subsubsection{\label{sec:ChiScaling}\Gls{SEF} scaling with bond dimension}
\begin{figure}[tb]
	\centering
	\includegraphics[width=.475\textwidth]{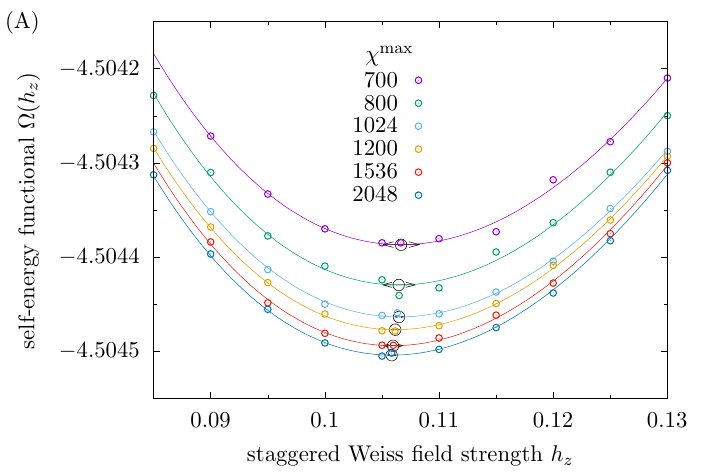}
	\includegraphics[width=.475\textwidth]{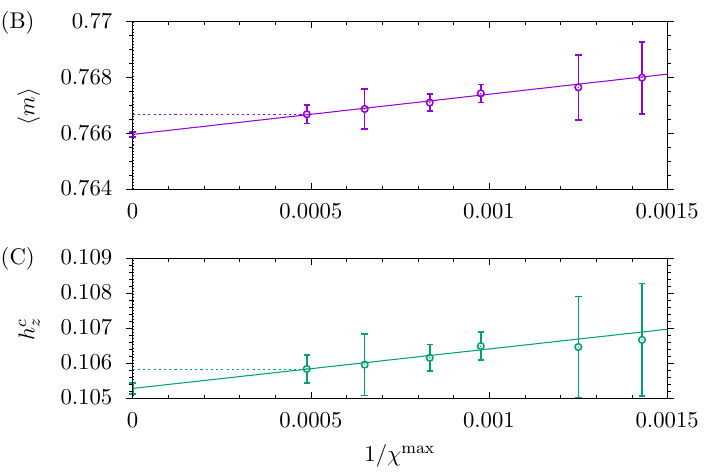}
	\caption
	{
		\label{fig:Chi_scaling}
		Scaling of the \gls{SEF} $\Omega$ in the maximal \gls{MPS} bond dimension $\chi^{\mathrm{max}}$ for the Betts cluster 22h5.
		Around the stationary point $\Omega(h_z^c)$ the functional can be fitted as a function of the Weiss field, shown as a line (A).
		The corresponding minima are indicated by black circles, the standard error of the minimum is indicated by black arrows.
		For different $\chi^{\mathrm{max}}$, the \gls{SEF} shows similar functional dependence on $h_z$.
		The staggered magnetization $\langle m\rangle$ on the lattice (B) and the Weiss field strength $h_z^c$ (C) are then scaled as a function of inverse bond dimension $1/\chi^{\mathrm{max}}$.
		We indicate both the standard deviation of the $1/\chi^{\mathrm{max}}$ extrapolated staggered magnetization as well as the value at the largest $\chi^{\mathrm{max}}$ (dashed lines).
	}
\end{figure}

In \cref{fig:Chi_scaling}, the value of the Weiss field $h_z^c$ as well as the staggered magnetization on the lattice, $\langle m\rangle$, are scaled in the inverse bond dimension.
It should be noted that other error bounds like the truncation error, which are well established for \gls{DMRG} (groundstate) calculations, are not applicable here since the calculation of the \gls{SEF} using \gls{MPS} is not following an overall variational principle.
For instance, the \gls{MPO} application within the band Lanczos algorithm only allows to estimate relative errors, and the propagation of these errors for the calculation of the \gls{SEF} is not clear.
Extrapolating the \gls{SEF} in the inverse bond dimension therefore seems to be the most appropriate approach.
The staggered magnetization scales linear in $1/\chi^{\mathrm{max}}$ and gives a lower estimate of the expectation value for the $\chi^{\mathrm{max}}\rightarrow\infty$ limit.
Since at some critical $\chi^c$ the \gls{MPS} representation is exact, we expect a constant value of $\langle m\rangle$ for $\chi^{\mathrm{max}}>\chi^c$.
The value of $\langle m\rangle$ at the largest calculated $\chi^{\mathrm{max}}$ therefore sets an upper bound for the staggered magnetization and can be used as a conservative upper error estimator.
However, the dependence of $h_z^c$ and $\langle m\rangle$ on $1/\chi$ is mild ($\mathcal{O}(10^{-3})$) and of the same order as the estimated error such that it can be neglected in practice for most applications as long as a sufficiently large bond dimension $\chi$ is used.
In the case of the shown 22-site Betts cluster, this would be the case for $\chi^{\mathrm{max}}\gtrsim 1000$.
Given the large computational cost of evaluating the cluster Green's function via the~\gls{MPS}\hyp based band Lanczos solver, systematically scaling expectation values in the \gls{MPS} bond dimension should remain the exception.
Instead, it offers the possibility to calculate spectral functions or observables for large clusters (e.g. using \gls{CPT}), for which symmetry\hyp breaking terms can be included via \gls{VCA}.
The rather mild dependence of the functional form of the \gls{SEF} and the related expectation values on the chosen \gls{MPS} bond dimension is important in this context.
\subsubsection{Spectral functions}
\begin{figure}[tb]
	\centering
	\includegraphics[width=1.\linewidth]{\imgpath/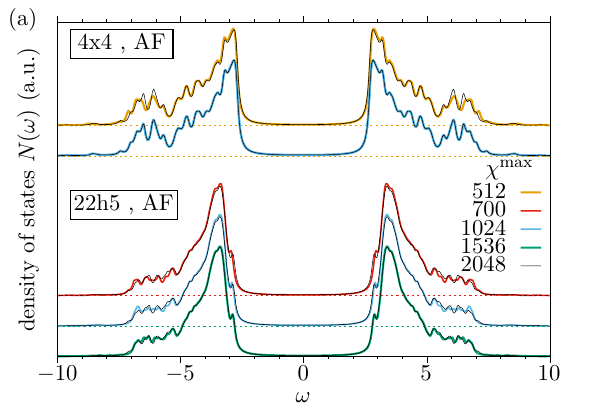}
	\includegraphics[width=1.\linewidth]{\imgpath/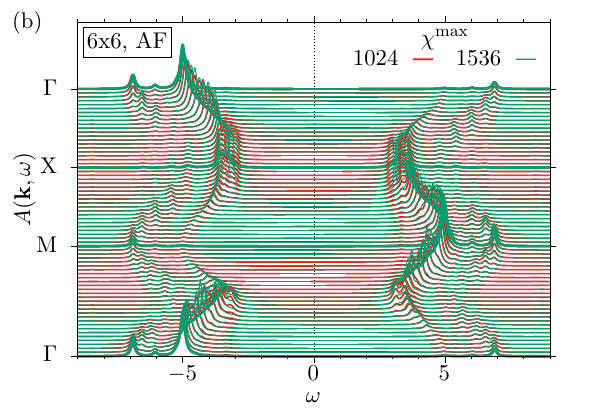}
	\caption
	{
		\label{fig:Akw}
		(a) \Gls{DOS} $N(\omega)$ of the $4\times4$ (top) and 22h5 (bottom) clusters for different maximal \gls{MPS} bond dimension $\chi^{\mathrm{max}}$.
		(b) shows the $\mathbf{k}-$resolved spectral function $A(\mathbf{k},\omega)$ of a $6\times6$ cluster for two different maximal bond dimensions.
		All calculations were done in the \gls{AFM} ordered state, i.e., at the stationary point of the \gls{SEF} for the respective clusters.
	}
\end{figure}
The spectral function defined via $A(\mathbf k,\omega) = -\frac{1}{\pi} \Im \mathbf G$ is commonly used to quantify the low\hyp energy effective single\hyp particle excitations, or more formally, it is the probability density to create a single\hyp particle excitation with energy $\omega$ and momentum $\mathbf k$.
In particular the single\hyp particle dispersion relation $\varepsilon(\mathbf k)$ and the \gls{DOS} $N(\omega) = \sum_{\mathbf k} \delta(\omega - \varepsilon(\mathbf k))$ can be directly derived from $A(\mathbf k, \omega)$.
Since it can be measured experimentally, for instance using \gls{ARPES}~\cite{hufner,RevModPhys.75.473}, it constitutes an important quantity to establish effective model descriptions for real materials.
As is usually done, we shift the Fermi momentum choosing $\varepsilon_F = 0$ such that a vanishing spectral function in a finite region around $\omega=0$ indicates the existance of a gap.
Next, we discuss the mild influence of the maximal \gls{MPS} bond dimension $\chi^{\mathrm{max}}$ on the spectral function.
In \cref{fig:Akw} we show the \gls{DOS} (A) and the spectral weight $A(\mathbf{k},\omega)$ along a high\hyp symmetry $\mathbf{k}$\hyp path in the Brillouin zone (B).
Since the dispersion of the model with nearest\hyp neighbor hopping is particle\hyp hole symmetric, we observe this symmetry also in the \gls{DOS}.
The chemical potential is located in the middle of the Mott gap, the system is hence Mott insulating as expected for the half\hyp filled Hubbard model in the strong\hyp coupling regime.
For comparatively small clusters, the spectral function $A(\mathbf{k},\omega)$ and the \gls{DOS} $N(\omega)$ can be converged in $\chi^{\mathrm{max}}$.
We illustrate this in the case of  the $4\times4$ cluster for $N(\omega)$ in \cref{fig:Akw}(a). 
The results obtained from restricting the bond dimension to $\chi^{\mathrm{max}}=512$ show already good agreement for excitation energies close to the gap, but the \gls{DOS} still shows differences at larger excitation energies around $\omega\sim \pm 6$.
For $\chi^{\mathrm{max}}\gtrsim1024$, the \gls{DOS} and spectral function of the $4\times4$ cluster are converged and do not change any more when increasing the bond dimension.
We found this convergence of the Green's function (or the \gls{DOS}), i.e., increasing the \gls{MPS} bond dimension to better capture the high\hyp frequency part, systematically throughout our simulations.
It can be understood by noting that in order to faithfully describe the large\hyp $\omega$ regime one has to be able to describe higher excited states with high accuracy, which requires large Krylov subspace dimensions.
However, constructing good approximations to Krylov vectors requires numerically well converged applications of $\hat H$ to previous Krylov vectors, a task that becomes increasingly complicated when it comes to the computation of higher\hyp order Krylov vectors.
In particular, the restricted bond dimension severely limits the approximation quality, which yields a degrading description of the high\hyp energy behavior, when decreasing the \gls{MPS} bond dimension.
Unfortunately, the discarded weight, which typically constitutes a more explicit error measure in the context of \gls{MPS} algorithms, can not be used here.
This is due to the fact that when variationally evaluating the \gls{MPO}\hyp \gls{MPS} application, i.e., minimizing the costfunction $C = \lVert \hat H \ket{\psi} - \ket{\varphi}\rVert^2$ over all \gls{MPS} representations $\ket{\varphi}$ with a maximum bond dimension $\chi$, the absolute error $C$ is unknown and only relative convergence within the sequence of site\hyp tensor updates can be achieved \cite{Paeckel2019}.
As a consequence, it is not possible to relate the total error incurred due to a finite bond dimension, i.e., the discarded weight to the error of the \gls{MPO}\hyp \gls{MPS} application.
For larger clusters like the $22h5$ cluster discussed previously in \cref{fig:Chi_scaling}, scaling $h_z^c$ and $\langle m\rangle$ still leads to minor changes when increasing the bond dimension up to $\chi^{\mathrm{max}}=2048$.
However, tracking the changes of the \gls{DOS} as a function of $\chi^{\mathrm{max}}$ reveals that only small changes close to the outmost band edges far from the gap are visible.
The \gls{DOS} for excitation energies $\vert\omega\vert\lesssim5$ is already converged for $\chi^{\mathrm{max}}=1536$ and even for $\chi^{\mathrm{max}}=1024$ the deviations from the converged solution are for most purposes negligible.
In \cref{fig:Akw}(b) we finally show the $\mathbf{k}$\hyp resolved spectral function $A(\mathbf{k},\omega)$ along the high\hyp symmetry path $(0,0)-(\pi,0)-(\pi,\pi)-(0,0)$ for one of the largest clusters we systematically studied, the $6\times6$ cluster.
The spectral function was calculated from the \gls{CPT} Green's function, \cref{eq:CPT}, using the periodization scheme proposed in Ref.~\onlinecite{Senechal2000}.
In contrast to small and intermediate clusters, the \gls{SEF} of the $6\times6$ cluster still shows notable differences as a function of the \gls{MPS} bond dimension up to $\chi^{\mathrm{max}}=2048$.
They translate to the spectral function in form of small differences in the low\hyp energy excitations, most visible around the X point, $\mathbf{k}=(\pi, 0)$.
Nevertheless, the most salient features of the spectral function are already converged in the bond dimension for a remarkably small $\chi^{\mathrm{max}}=1536$.
This can be seen as an advantageous feature of the \gls{MPS} solver, which enforces due to the energy truncation scheme first the convergence of poles of the Green's function that contribute most spectral weight.
The present \gls{MPS} solver therefore even allows to interpret spectral functions of large clusters for which a full convergence in terms of the \gls{MPS} bond dimension cannot be achieved.
\subsubsection{\label{Sec:FSscaling}Finite size scaling}
\begin{figure}[tb]
	\centering
	\includegraphics[width=.5\textwidth]{\imgpath/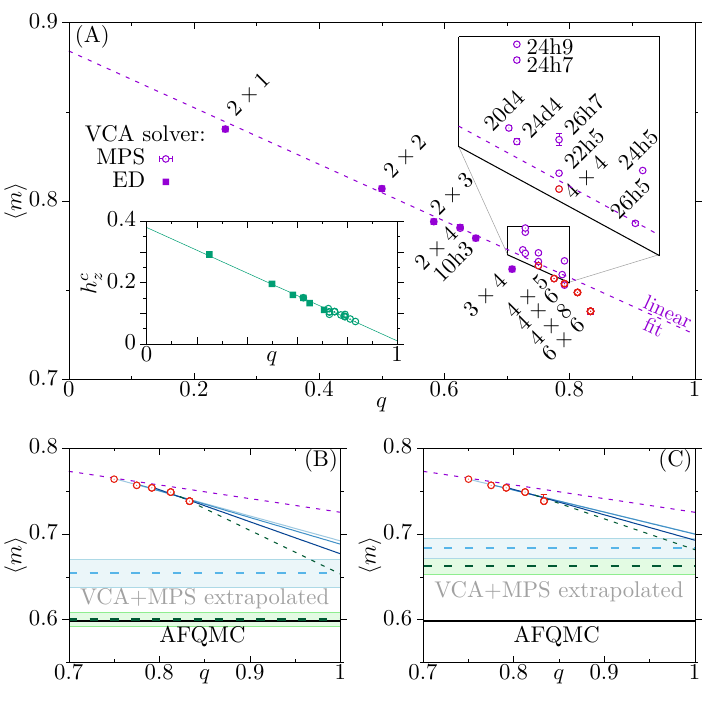}
	\caption
	{
		\label{fig:FS_scaling}
		Finite\hyp size scaling of \gls{VCA} data for different rectangular and Betts clusters.
		(A) The staggered magnetization $\langle m\rangle$ and the magnetic Weiss field on the cluster $h_z$ (inset) are scaled as a function of the parameter $q$ described in text.
		The linear fit (violet dashed line) takes into account all data points, the band Lanczos results for large rectangular clusters (highlighted in red) are used in panels (B,C).
		(B) Extrapolation using a nonlinear ansatz as described in the text, based on the error estimate from the $1/\chi^{\mathrm{max}}$ extrapolation.
		The extrapolated staggered magnetization including (excluding) the last interpolation is shown as dashed green (blue) line, see text.
		We indicate the $1\sigma$ confidence interval of the extrapolation in light green (blue).
		(C) Same as panel (B), but with $m(\chi^{\mathrm{max}})$ as a conservative upper error estimate.
		In all cases, the linear extrapolation is improved with respect to the exact \glsentrylong{AFQMC} result, taken from Ref.~\cite{Seki2019}.
	}
\end{figure}
In order to perform a finite\hyp size scaling of the staggered magnetization, calculations need to be performed using different cluster sizes.
However, even for fixed cluster size, the cluster geometry has an impact on the \gls{SEF} since the corresponding cluster Green's functions include long\hyp range correlations to a different extent.
To include this information for the finite\hyp size scaling, we use the quality factor introduced by \textit{D. S\'en\'echal} in Ref.~\onlinecite{Senechal2008}.
It amounts to comparing the intra\hyp cluster hopping terms $N_{t}^{cl}$ to the number of hopping terms per site in the infinite lattice and therefore takes into account both the cluster size and geometry.
In the square lattice, this amounts to $q=N^{cl}_t/2L$.
The scaling factor $q$ thereby contrasts with other commonly used scaling parameters like $1-1/L$, which include only information of the cluster size.
When using this quality factor, the extrapolation of the Weiss field strength $h^c_z$ to the thermodynamic limit leads to a vanishing field strength whereas the expectation value of the staggered magnetization on the lattice $\langle m\rangle$ extrapolates to a finite value, see \cref{fig:FS_scaling}(A).
In the infinite\hyp cluster\hyp size limit, $q=1$, the symmetry\hyp breaking field is not needed since the symmetry is  spontaneously broken.
There, the value of the staggered magnetization $\left.\langle m\rangle\right\vert_{q=1}$ simply corresponds to the order parameter on the lattice. %
At half\hyp filling, $\langle m \rangle$ can be calculated numerically exactly using \gls{AFQMC} \cite{AFQMC_0,AFQMC_1,AFQMC_2,AFQMC_3,AFQMC_4,Becca2017}.
Comparing to the \gls{AFQMC} result (dashed black line) of \textit{Seki and Sorella} \cite{Seki2019} we see that the linear scaling in $q$ (dashed, violet graph) would lead to an overestimation by $\sim 20\%$ of the staggered magnetization, i.e., the linear dependence $\langle m\rangle$ on $q$ seems to hold only for small and intermediate cluster sizes.
Indeed, for the largest clusters calculated with our \gls{MPS} band Lanczos solver, we already observe a significant deviation from linear scaling (cf. the red highlighted symbols in \cref{fig:FS_scaling}).
We believe that besides others, the nonlinear $q$\hyp dependence is generated from the fact that in the presented calculations we did not vary the hopping amplitudes $t_b$ at the cluster boundaries.
In that case, an extrapolation would still be possible even though the exact functional dependency is unknown, and we propose a modified finite\hyp size extrapolation procedure, in the following.
Even though in our \gls{VCA} computations we fixed potential variational parameters such as $t_b$, an extrapolation in the cluster size is still possible if we take into account that the variational parameters exhibit different functional dependencies with respect to the quality factor.
For the antiferromagnetic Weiss field strength $h^c_z$, which mimics the effect of long\hyp range order in the thermodynamic limit, a functional dependency of the form $h^c_z \sim 1-L^{-1}$ is expected.
This exactly corresponds to a scaling of the quality factor and is verified by our results as shown in the inset of \cref{fig:FS_scaling}.
On the other hand, for the hopping amplitudes at the cluster boundaries a functional dependency $t - t_b \sim 1-L^{-1/2}$ could be expected, since the number of boundary sites of the square lattice scales as $L^{1/2}$.
Assuming that the corrections $t-t_b$ vanish as $1-L^{-1/2}$, these will dominate the magnetization for small $q$.
Thereby, we can estimate the effect of fixing $t_b = t$ in our \gls{VCA} computations by a rescaling of the otherwise linear dependency of the magnetization in $q$.
We attempt to account for this nonlinear behavior by: (1) Linearly extrapolating $\braket{m}(q\rightarrow 1,q_0)$ for the largest rectangular clusters (cf. the red highlighted symbols in \cref{fig:FS_scaling}) on a domain $[q_0,1]$ where we varied $q_0\in [0.775,0.8125]$.
The obtained linear extrapolations are indicated by the blue and dashed green lines in \cref{fig:FS_scaling}(B,C).
(2) We then make the simplest nonlinear ansatz for the functional dependency on $q_0$
\begin{align}
	\braket{m}(q\rightarrow 1,q_0) = \alpha (q_0-1)^2 + m_\infty \;.
\end{align}
This ansatz has a stationary point at $q_0=1$, resembling the expected asymptotic behavior of the magnetization when scaling the number of cluster sites to the thermodynamic limit.
We extrapolate the data with (without) the last linear extrapolation ($q_0=0.8125$, green dashed line) and obtain the results shown in \cref{fig:FS_scaling} as horizontal green (blue) dashed line.
The corresponding $1\sigma$ confidence interval of the extrapolation is indicated by the shaded area.
When including the $q_0=0.8125$ extrapolation, the extrapolated value $m_{\infty}=0.602\pm0.008$ agrees with the \gls{AFQMC} result $m^\mathrm{QMC}_\infty = 0.5982$ within the confidence interval. 
A more conservative estimation of the extrapolation error can be obtained by taking the value $\langle m\rangle\vert_{\chi^{\mathrm{max}}=\mathrm{max}}$ as an upper error bound, see \cref{fig:FS_scaling}(C).
In this case, the extrapolated magnetization $m_{\infty}=0.663\pm0.010$ overestimates the exact value by $11\%$, but still improves a linear\hyp in\hyp $q$ fitting of all data.
We therefore find that with the cluster sizes accessible via the \gls{VCA}+\gls{MPS} ansatz, it is possible to estimate values for local observables in the thermodynamic limit to a very high precision, even without exploring all possible variational parameters within \gls{VCA}.
As a final remark, we note that the constructed fitting procedure is only based on the fact that corrections in the order parameter generated by fixing variational parameters can exhibit different functional dependencies when scaling the system size.
Thus, while this approach is generally applicable, it cannot establish absolute error bars, as long as the explicit extrapolation functions are unknown.
\subsubsection{Application to hole-doped CaCuO$_2$}
To illustrate the applicability of the \gls{MPS}\hyp based band Lanczos within \gls{VCA} in situations where no (numerically) exact solutions exist, we employ it to the single-band Hubbard model away from half\hyp filling.
Motivated by \textit{ab initio} low\hyp energy parametrizations of the ``infinite\hyp layer'' cuprate CaCuO$_2$ at different levels of hole\hyp doping \cite{Moree22} we study here the $t-t^{\prime}-U$ Hubbard model at densities between $n=1$ and $n=0.9$.
CaCuO$_2$ has been in the focus over the last years, since it is isostructural to the nickelate NdNiO$_2$.
The latter was recently synthesized hole\hyp doped and found to show emergent superconductivity at low temperature \cite{Nickelates2}.
Here, we slightly simplify the model given in Ref.~\onlinecite{Moree22} and use the parametrizations listed in \cref{Tab:modelparam}:
We focus for undoped and hole\hyp doped CaCuO$_2$ with $0\leq \delta \leq 10\%$ on the parametrization for $\delta=0$, at $10\%$ doping we compare to the parametrization derived for $\delta=0.1$.
Longer\hyp ranged hopping terms are rather small, which is why we focus here on the nearest\hyp {} and next\hyp nearest\hyp neighbor hopping terms only, thereby neglecting hopping terms with $\vert t_i/t\vert<0.11$.
Furthermore, the nonlocal interaction terms $V_n/U<0.25$ derived in Ref.~\onlinecite{Moree22} are not included here, since they would require a mean\hyp field decoupling at the cluster boundaries \cite{Aichhorn2004}.
\begin{table}[b!]
	\caption
	{
		\label{Tab:modelparam}
		Parameters of the low\hyp energy effective models of $\delta=0,10\%$ hole-doped CaCuO$_2$, taken from Ref.~\onlinecite{Moree22}.
	}
	\begin{center}
		\begin{tabular}{l|c|c|c|}
			$\delta$&$t$ (eV)&$t^{\prime}/t$&$U/\vert t\vert$\\\hline
			0&-0.521&-0.21&8.60\\
			0.1&-0.521&-0.25&8.10\\
		\end{tabular}
	\end{center}
\end{table}
The full model has been studied recently with respect to its superconducting and \gls{AFM} phases by employing a many\hyp variable variational Monte Carlo technique based on variational wave functions on large square clusters \cite{Schmid23}.
However, the study did not comprise the calculation of quantities that rely on Green's functions such as the spectral function.
Using a slightly different \textit{ab initio} one\hyp band model, the \gls{AFM} phase of CaCuO$_2$ has also been studied by single\hyp site \gls{DMFT} \cite{Karp2020} as well as with its cluster extension \cite{Karp2022}, which allows us to compare our results to techniques that include dynamical and -- in case of \gls{CDMFT} -- also short\hyp range spatial fluctuations.
Here, we focus on the \gls{AFM} phase at and off half\hyp filling, thereby limiting the Weiss fields within \gls{VCA} to a staggered magnetic field $M_c$ and -- in case of doping -- to the cluster chemical potential $\mu^{\prime}$.
We furthermore choose a chemical potential $\mu$ that optimizes the free energy functional, thereby allowing for a thermodynamically stable electron density $n$ \cite{Aichhorn2006}.
In order to allow for commensurate electron fillings on the cluster at $\delta=0.1$, we choose a 20\hyp site cluster ($4\times5$ or 20p0, see \cref{fig:App_4x5}).
\begin{figure}[tb]
	\centering
	\includegraphics[width=.5\textwidth]{\imgpath/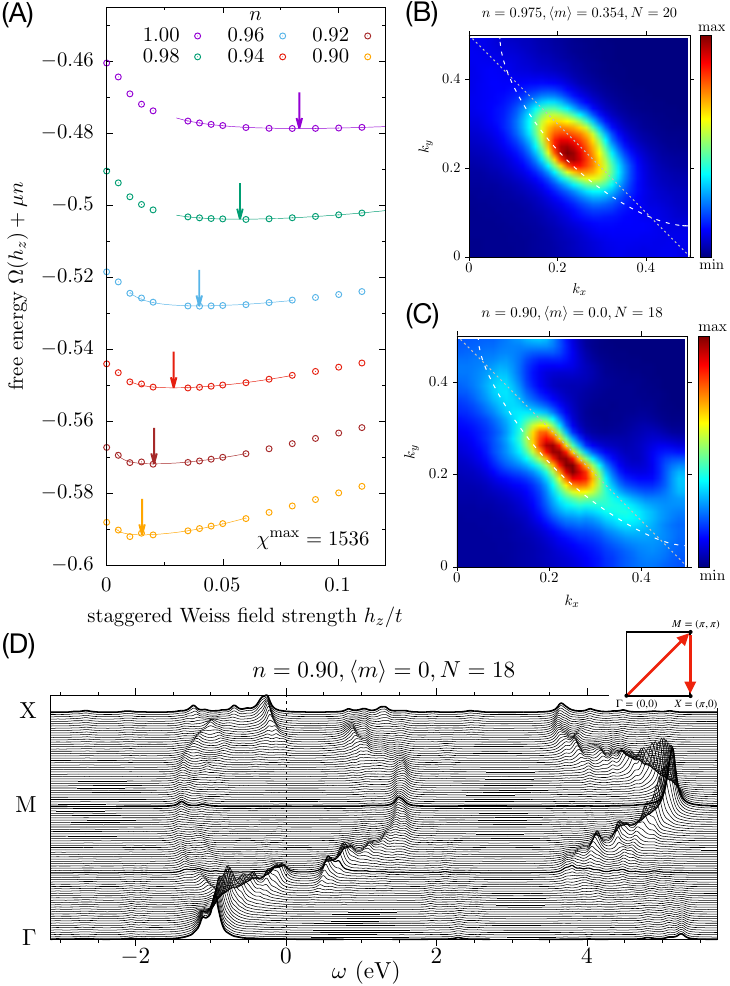}
	\caption
	{
		\label{fig:CC11}
		\Gls{VCA} simulation of hole-doped and pristine CaCuO$_2$.
		(A) The \gls{SEF} $\Omega(h_z) + \mu n$ with respect to the staggered Weiss field for different electronic densities $n$, limiting the quantum number sector on the cluster to $N=20$ electrons.
		Lines indicate a fit around the minima, which are marked by arrows.
		The \gls{AFM} solution can be stabilised for $1\leq n\lesssim 0.9$.
		(B) First quadrant of the many\hyp body Fermi surface in the \gls{AFM} state at $n=0.975$ (and $N=20$) and in the \glsfirst{PM} state (C) at $n=0.90$ for the $N=18$ electron quantum sector.
		Dashed white lines indicate the non-interacting Fermi surface and the \gls{AFM} BZ boundary is shown as a dashed grey line.
		(D) The single\hyp particle spectral weight $A(\mathbf{k},\omega)$ of the \gls{PM} $\delta=10\%$ doped system is shown for wave vectors along the high\hyp symmetry directions shown in the bottom panel.
	}
\end{figure}
In \cref{fig:CC11}(A) we show the evolution of the \gls{SEF} upon hole\hyp doping. 
For the undoped system ($n=1.0$), the \gls{SEF} shows a minimum as a function of the Weiss field strength at $h_z\sim 0.08 t$, which corresponds to an \gls{AFM} insulator with staggered magnetization $\langle m\rangle=0.377$.
The value of the magnetization is in good agreement with the one reported in Ref.~\onlinecite{Karp2022} from \gls{CDMFT} ($\langle m\rangle=-0.371$).
This consistency could be expected since the correlations of an undoped Mott insulator in the strong\hyp coupling region are short ranged such that both the $2\times2$ cluster and our $20$\hyp site cluster capture the correlations correctly.
Upon doping, the minimum of the functional shifts continuously to smaller values of $h_z$ and the condensation energy $F_{\text{PM}}-F_{\text{AFM}}$ reduces, finally leading to a transition to the nonmagnetic phase at $n\approx0.9$.
Comparing to the end point of the \gls{AFM} solution in the finite\hyp temperature phase diagram obtained within \gls{CDMFT}, we again find good agreement.
Not surprisingly, the absence of dynamical and thermal fluctuations leads to a larger staggered magnetization close to the phase transition within \gls{VCA}.
Having established the correct description of the \gls{AFM} phase, we focus next on the spectral function at selected densities to exploit the good spatial resolution within our cluster approach.
Panels (B) and (C) of \cref{fig:CC11} show the Fermi surface at $\delta=2.5\%$ and $\delta=10\%$ doping respectively as obtained from the periodized many\hyp body Green's function, i.e., the spectral weight $A(\mathbf{k},\omega)$ at $\omega=0$.
At low doping, the intensity close to the $(\pi,0)$ and $(0,\pi)$ points disappears and only spectral weight around $\mathbf{k}=(\pi/2,\pi/2)$ is left.
The absence of spectral weight at the nodal points while preserving spectral weight at the antinodal point is characteristic for hole\hyp doped cuprates \cite{Marshall1996,Loeser1996,Ding1996}.
Here, we find a pronounced feature on the not\hyp backfolded part of the spectrum, revealing Fermi arcs in contrast to Fermi pockets.
Upon larger doping, non\hyp zero spectral weight appears close to the non\hyp interacting dispersion (panel C), finally showing a full Fermi sheet.
These results are in good qualitative agreement with the Fermi surface obtained within \gls{CDMFT} on a $2\times2$ cluster (see Fig.~3(e),(f) of Ref.~\cite{Karp2022}), even though the differences in cluster size lead to a richer structure of the Fermi surface at $\delta=10\%$ within the \gls{VCA} calculation on a 20\hyp site cluster.
The larger spectral weight at the antinodal point and reduced weight along the remainder of the Fermi sheet are reminiscent of the strong\hyp coupling pseudogap.
This can be seen in panel D, where the spectral function $A(\mathbf{k},\omega)$ is shown along the high\hyp symmetry $\mathbf{k}$\hyp path $\Gamma-M-X$ for the $10\%$\hyp doped paramagnetic system:
The Mott gap being located at positive energies $\sim1.6\ \mathrm{eV} < \omega \sim 3.6\ \mathrm{eV}$ as expected for electron\hyp doped Mott insulators \cite{Meinders1993}, the qualitative difference of the spectral weight at the Fermi level $\omega=0$ along $\Gamma-M$ and $X-M$ is visible.
In particular, at the antinodal point the peak of the spectral intensity does not coincide with the Fermi level, but a quasiparticle peak at $\sim -30\ \mathrm{meV}$ leads to the spectral weight at $E_F$.
Here, no long\hyp range \gls{AFM} order is present ($h_z=0$), but the 20\hyp site cluster captures short\hyp range \gls{AFM} fluctuations, which are known to cause the pseudogap formation in the strong\hyp coupling regime of the Hubbard model \cite{Senechal2004,Simkovic2024}.
Finally, we note that the description of the $10\%$-doped system within \gls{VCA} is best done with a cluster that has a commensurate electron filling (i.e., $N=18$ electrons on the 20\hyp site cluster).
In contrast to the half\hyp filled cluster, this electronic configuration corresponding to a correlated metal with ungapped energy spectrum results in a higher bipartite entanglement entropy of the groundstate wavefunction, see \cref{fig:App_4x5} in the appendix.
Whereas the presence of a finite staggered Weiss field $h_z$ helps to reduce the bipartite entanglement entropy at half filling, the reduction is much smaller for the cluster with $N=18$ electrons. 
Its description with \gls{MPS} is therefore more challenging than the half\hyp filled cluster.
This is contrary to \gls{ED}, where the dimension of the Hilbert space is the leading criterion for the numerical complexity, resulting in a numerically more challenging situation at half filling.
\section{\label{sec:conclusion}Conclusions and Outlook}
We introduce a band Lanczos solver based on \gls{MPS} for quantum cluster methods such as the \gls{VCA}.
While a na\"ive implementation of \gls{MPS} as cluster solver does not lead to a substantial improvement over using \gls{ED} techniques, we showed that significantly larger cluster sizes can be treated with high accuracy by combining the \gls{MPS}+band Lanczos solver with a suitable energy truncation, controlling the loss of orthogonality of the Lanczos states and introducing a robust convergence criterion.
We demonstrate the potential of the approach by computing the staggered magnetization $\langle m \rangle$ of a single\hyp band Hubbard model on a \gls{2D} square lattice.
Treating different cluster sizes and shapes (regular rectangular as well as Betts clusters), we extrapolate our \gls{MPS} data in the bond dimension $\chi$, which we show to yield a systematic improvement of the approximation quality of the Green's function and its derived quantities.
In this way we can exploit the increased number of accessible cluster sizes and geometries to extrapolate observables such as the staggered magnetization into the thermodynamic limit.
We show that at half filling and for intermediate interaction strengths $U/t = 8$ (i.e., in the Mott insulating regime, where the interaction strength is of the order of the bandwidth of the noninteracting system) an anomalous finite\hyp size scaling with a crossover length scale is observed.
Only for rectangular cluster sizes that are large enough, i.e., $4\times 4$ and larger, a deviation from an otherwise linear scaling is obtained, which can be extrapolated towards the thermodynamic limit.
The resulting value of $\langle m \rangle$ in the thermodynamic limit is improved over a simple linear\hyp in\hyp $q$ extrapolation and can -- depending on the extrapolation scheme -- even agree within error margins with numerically exact \gls{QMC} data. 
In contrast, with the smaller cluster sizes amenable to \gls{ED}, the extrapolation in cluster size leads to a value, which overestimates the \gls{QMC} result by $\sim 20\%$.
We illustrate the applicability of the approach to more complex situations by studying the $t-t^{\prime}-U$ Hubbard model at and off half filling as relevant to cuprate materials such as electron\hyp doped CaCuO$_2$.
The \gls{MPS}\hyp \gls{VCA} ansatz allows to study clusters with commensurate electron densities even at low electron doping.
Given the significant increase of the number of cluster sites, we envisage that our approach has the potential to substantially improve the investigations in multi\hyp band Hubbard\hyp like models.
Here, Hund's coupling leads to an interplay between orbital, spin, and electronic degrees of freedom \cite{Georges2013,deMedici2017}, which affects material properties like superconductivity in nickelates and iron based superconductors, orbital selectivity in iron chalcogenides or optoelectronic and photovoltaic properties \cite{Lanata2013,Arribi2021,PRB2bandHM,CorrelationsRuddlesdenPopper,Nickelates1,Nickelates2,RuddlesdenPopperOptoElectronics,RuddlesdenPopperPV,RuddlesdenPopperOptoElectronics}.
Although the interactions are local in nature, the higher complexity of these \gls{2D} systems poses an even stronger challenge to their theoretical description.
While it is essentially impossible for \gls{ED}\hyp based solvers to treat cluster sizes large enough to perform meaningful studies on such multi\hyp orbital systems, the \gls{MPS}+band Lanczos approach makes it possible to treat also systems with larger local Hilbert spaces in a controlled way, so that it appears natural to extend our approach to such multi\hyp band situations.
\section*{Acknowledgments}
All \gls{MPS} band Lanczos simulations were performed with the \textsc{SymMPS} toolkit \cite{symmps}.
We thank the Grand Equipement National de Calcul Intensif for providing supercomputing time at IDRIS and TGCC (Project No. A0130912043). 
The authors gratefully acknowledge the computing time provided to them on the high performance computers noctua2 \cite{noctua2} at the NHR Center PC2. 
These are funded by the Federal Ministry of Education and Research and the state governments participating on the basis of the resolutions of the GWK for the national highperformance computing at universities (www.nhr-verein.de/unsere-partner).
We also acknowledge a mobility grant of the Franco-Bavarian University cooperation centre (BayFrance).
Furthermore funding through the ERC Starting Grant from the European Union's Horizon 2020 research and innovation programme under grant agreement No 758935 is greatfully acknowledged. 
S. P. acknowledges support by the Deutsche Forschungsgemeinschaft (DFG, German Research Foundation) under Germany’s Excellence Strategy-426 EXC-2111-390814868.
Support by the Deutsche Forschungsgemeinschaft (DFG, German Research Foundation) - 207383564/FOR1807 project P7 at an early stage of this work is acknowledged.
\bibliography{refs}%

\appendix%
\setcounter{equation}{0}%
\setcounter{figure}{0}%
\setcounter{table}{0}%
\setcounter{page}{1}%
\makeatletter%
\renewcommand{\theequation}{A\arabic{equation}}%
\renewcommand{\thefigure}{A\arabic{figure}}%
\numberwithin{equation}{section}
\section{\label{app:Green}$Q$-matrix formulation of the cluster Green's function}
To rewrite the cluster Green's function, we use the $\mathbf{Q}$\hyp matrix formulation of the Lehmann representation introduced in Ref.~\cite{Qmat}:
\begin{align}
	G^{\prime}_{\alpha,\beta}(\omega) = \sum_m\frac{Q^{\nodagger}_{\alpha,m\vphantom{\beta}}Q^{\dagger}_{m,\beta}}{\omega-\omega^{\prime}_m}\;,
\end{align}
with $\alpha,\beta$ being compound site and spin indices, and excited states $m$.
The $\mathbf{Q}$\hyp matrix is defined as the concatenation of the electron and hole matrices
\begin{align}
	Q^{\vphantom{(e)}}_{\alpha,m} =& \left( Q_{\alpha,r}^{(e)}, Q_{\alpha,s}^{(h)}\right)\,, \\
	Q_{\alpha,r}^{(e)} =& \braket{0 \vert \hat c^{\vphantom{(e)}}_{\alpha} \vert r}\,, \\
	Q_{\alpha,s}^{(h)} =& \braket{s \vert \hat c^{\vphantom{(e)}}_{\alpha} \vert 0} \;,
\end{align}
where $\ket{0}$ is the ground state with energy $E_0$.
The corresponding excitation energies needed in \cref{eq:sum2} read $\omega^{\prime}_m = \left(\omega^{(e)}_r,\omega^{(h)}_s\right)=\left(E^{\noprime}_r-E^{\noprime}_0,E^{\noprime}_0-E^{\noprime}_s\right)$.
To obtain the poles $\omega$ for \cref{eq:sum1} we rewrite the \gls{VCA} Green's function as
\begin{align}
 \mathbf{G}	&=\frac{1}{\mathbf{G}_0^{-1}-\Sigma} = \frac{1}{(\mathbf{G}^{\prime})^{-1}-\mathbf{V}}\\
 			&= \mathbf{Q}\frac{1}{\mathbf{g}^{-1}-\mathbf{Q}^{\dagger}\mathbf{V}\mathbf{Q}}\mathbf{Q}^{\dagger}\;,
			\label{eq:end}
\end{align}
where $\mathbf{g}^{-1} = \omega - \mathbf{\Lambda}$ with $\Lambda^{\noprime}_{m,l} = \delta^{\noprime}_{m,l} \omega^{\prime}_{m\vphantom{l}}$ so that the poles $\omega_m$ are nothing but the eigenvalues of $\mathbf{\Lambda}+\mathbf{Q}^{\dagger}\mathbf{V}\mathbf{Q}$.
The $\mathbf{Q}$\hyp matrix and the one\hyp particle excitations $\omega^{\prime}$ are calculated via the band Lanczos algorithm.
\numberwithin{equation}{section}
\section{\label{app:hl}Generalized residual}
The generalized residual $\delta R^D(\tau)$ can be computed at no additional cost using the approximation of the exponential in the Krylov subspace
\begin{align}
	\delta R^D(\tau) = \beta^D \braket{f^{D-1}| \mathrm e^{-\mathrm i \hat H^D \tau} | f^D} \;. \label{app:eq:generailzed-residual}
\end{align}
While this provides a reliable bound for the usual Lanczos procedure,~\cref{app:eq:generailzed-residual} in the context of~\cref{eq:hl} is exact for the case of the band Lanczos only, if the Krylov dimension $D$ and the Krylov order $N$, i.e., the number of applications of $\hat H$ to the set of $L$ initial states, are related via $D = N\cdot L$.
The reason behind this limitation is that deriving the Hochbruck\hyp Lubich bound~\cref{eq:hl} for the generalized residual it is assumed that the residual Krylov vector $\ket{f^D}$ has a Krylov order $N+1$ while $\hat H^D$ is approximated with Krylov order $N$, only.
In princple this means that \cref{eq:hl} can only be applied after a whole sequence of $L$ candidate states has been constructed.
However, in practise we observed that also within such a sequence, the generalized residual $\delta R^D(\tau)$ yields satisfying error estimations with fluctuations being of an order of magnitude, at the most.

\section{\label{app:1D}1D mapping and entanglement entropy within MPS}
\begin{figure}[!h]
	\includegraphics[width=\columnwidth]{\imgpath/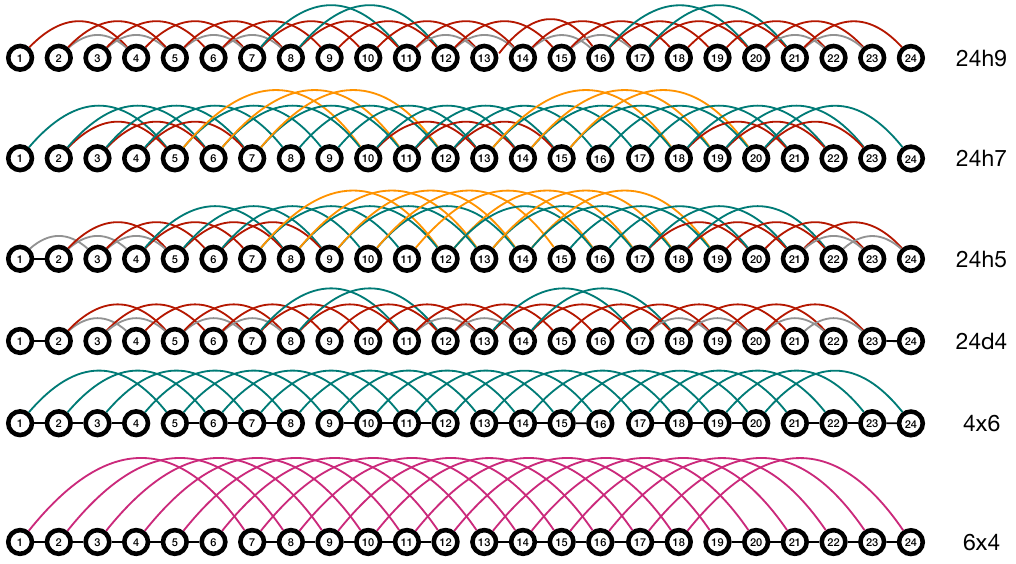}
	\caption
	{
		\label{fig:Apx1Dmap}
		\Gls{1D} mapping of the \gls{2D} $24$\hyp site clusters discussed in the main text. 
		The range of the hopping terms $t_{ij}$ is illustrated by the height of the respective lines connecting two sites $i$ and $j$.
	}
\end{figure}
\begin{figure}[!h]
	\includegraphics[width=\columnwidth]{\imgpath/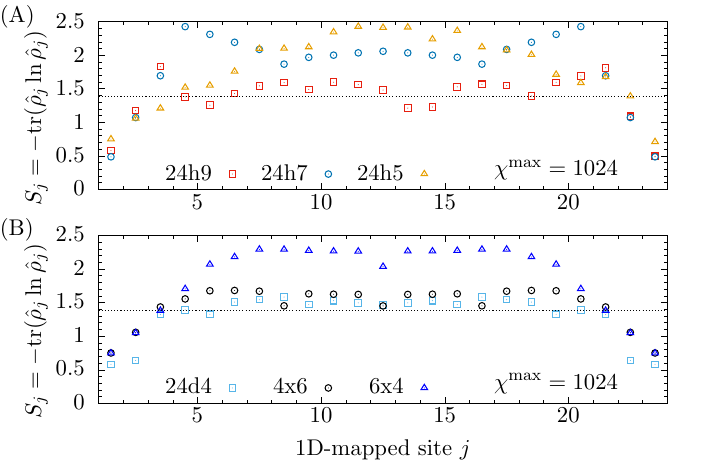}
	\caption
	{
		\label{fig:App_S_24site}
		The bipartite entanglement entropy $S_j=-\mathrm{tr}(\hat\rho_j\ln\hat\rho_j)$ of the groundstate wave function of the $24$\hyp site clusters depicted in \cref{fig:Apx1Dmap} for $\chi^{\mathrm{max}}=1024$ at a Weiss field strength of $h_z=0.10$.
	}
\end{figure}
Efficiently representing the Lanczos vectors in terms of \gls{MPS} requires a certain care when choosing the mapping of the cluster geometry to a \gls{1D} chain.
In particular, the computational efficiency and therefore also the achieved numerical accuracy given a fixed bond dimension $\chi$ crucially depends on the coupling ranges.
This can be understood directly by noting that using \gls{MPS}, two\hyp point correlation functions are approximated by a sum of $\chi^2$ exponentials \cite{Schollwoeck201196}.
Therefore, in choosing a specific chain mapping one should always try to minimize the coupling ranges in order to render correlations as short\hyp ranged as possible to achieve a high approximation quality of the correlation functions.
In \cref{fig:Apx1Dmap}, we show chain mappings for the various clusters computed in the main text and indicate the resulting couplings between the lattice sites of the chain.
\begin{figure}[!t]
	\includegraphics[width=\columnwidth]{\imgpath/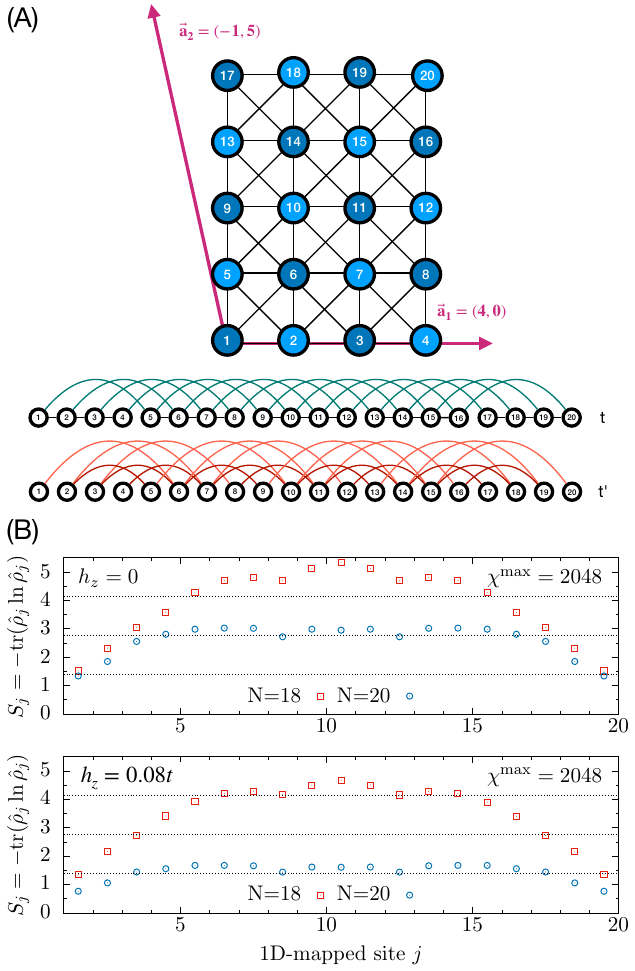}
	\caption
	{
		\label{fig:App_4x5}
		(A) Cluster 20p0 used for the $t-t^{\prime}-U$ Hubbard model (top) as well as its \gls{1D} mapping of the (next\hyp{})nearest\hyp neighbour hopping terms ($t^{\prime}$) $t$ (bottom).
		(B) Bipartite entanglement entropy $S_j$ of the groundstate wave function at Weiss field strengths of $h_z=0$ (\gls{PM}) and $h_z=0.08t$ (\gls{AFM} at half filling) for $N=18$ and $N=20$ electrons on the 20p0 cluster.
		The maximal \gls{MPS} bond dimension is $\chi^{\mathrm{max}}=2048$ and the dashed lines are a guide to the eye.
	}
\end{figure}
The bipartite entanglement entropy $S_j = \operatorname{tr} \hat\rho_j \ln \hat\rho_j$ constitutes a viable measure to quantify the range of correlations encoded in the~\gls{MPS}, where $\hat\rho_j$ denotes the reduced density matrix when introducing a bipartition between sites $(j,j+1)$.
For instance, considering a pair of maximally entangled (correlated) lattice sites the bipartite entanglement reaches its maximum value $S_0=\ln d$ where $d$ denotes the dimension of the local Hilbert spaces.
For the case of a rectangular lattice with $L_x\times L_y$ sites mapped to a one\hyp dimensional chain, one typically introduces a snake mapping, decomposing the system into either $L_x$ or $L_y$ coupled chains where the last site of the $n$th chain is connected to the first site of the $n+1$st chain.
As a consequence, a bipartition between two sites can exhibit maximum values $S_{nL_{x,y}} = L_{x,y}\ln d$.
It is therefore beneficial to decompose the grid in such a way that the length of the chains is minimized, i.e. for $L_x < L_y$ chains are formed along the $x$\hyp direction.
We demonstrate this effect by computing the ground state of a $24$\hyp site cluster choosing both types of chain mappings for a rectangular cluster ($4\times 6$ and $6\times 4$ sites where the first number denotes the length of the chains).
In \cref{fig:App_S_24site} we show the resulting bipartite entanglement entropies at the example of a maximum bond dimension $\chi=1024$, demonstrating that for chains with length $6$, the maximally observed entanglement entropy is larger by a factor of roughly $1.5$ compared to the case of chains with length $4$.
This illustrates how the size of the boundary determines the entanglement occuring in the \gls{MPS}, which, in this case, nicely reproduces the area law of entanglement for ground states.
For comparison, we also show bipartite entanglement entropies for various Betts clusters and we observe the largest values for the 24h5\hyp cluster, which is consistent with large number of long\hyp ranged couplings spanning $6$ lattice sites.

\end{document}